\documentclass[11pt,a4paper]{article}
\usepackage{jcappub}
\pdfoutput=1

\usepackage{ruths_commands}
\usepackage{martins_commands}

\begin{document}


\title{Large deviations for halos and voids:\\ beyond perturbative non-gaussianities}

\author[a]{Martin Teuscher,}
\author[b]{Ruth Durrer,}
\author[c]{Julien Grain,}
\author[a]{Killian Martineau,}
\author[a]{Aurélien Barrau}

\affiliation[a]{Laboratoire de Physique Subatomique et de Cosmologie, \\ 53 avenue des Martyrs, 38000 Grenoble, France}
\affiliation[b]{Département de Physique Théorique, Université de Genève,\\24 quai Ernest Ansermet, 1211 Genève 4, Switzerland}
\affiliation[c]{Université Paris-Saclay, CNRS, Institut d’Astrophysique Spatiale, 91405, Orsay, France}

\emailAdd{teuscher@lpsc.in2p3.fr}
\emailAdd{ruth.durrer@unige.ch}
\emailAdd{julien.grain@universite-paris-saclay.fr}
\emailAdd{martineau@lpsc.in2p3.fr}
\emailAdd{barrau@in2p3.fr}

\abstract{The excursion-set formalism provides a key connection between primordial density fluctuations and the abundance of cosmic structures such as dark matter halos and voids, traditionally assuming Gaussian random walks. In this work, we extend this framework to fluctuations whose distribution presents strongly non-Gaussian tails. Such tails are beyond the reach of perturbative approaches to primordial non-Gaussianity based on moment expansion. We address the problem with rigorous, analytical derivations relying on the large deviation principle, suited for the study of rare fluctuations. We derive new first-passage time distributions for random walks with non-Gaussian statistics and obtain updated predictions for the halo mass function.  We also study the two-barrier problem relevant to cosmic void formation, leading to a new analytical prediction for the void size function, with improved accuracy on  large scales. Our results demonstrate the potential of large deviation techniques as a bridge between inflationary scenarios, often leading to strongly non-Gaussian tails, and late-Universe observables.}


 \maketitle

\section{Introduction}
\label{sec:intro}

Primordial non-Gaussianities (PnG) constitute one of the primary targets of forthcoming cosmological surveys. While current observations of the cosmic microwave background (CMB) remain consistent with Gaussian primordial fluctuations \cite{Planck:2019kim_PNG_constr}, a large window of scales escape these constraints \cite{Beutler:2019ojk_Pk_out_of_CMB_scales},  and rare, large initial fluctuations are often predicted to obey non-Gaussian distributions \cite{Gow:2022jfb_nonpert_PBH_1pt_func,Pi:2022ysn_deltaN_pdf,Renaux-Petel:2015bja_review_png_after_planck2015,Tada:2016pmk_deltaN_bispectrum}.  The possibility of detecting even small departures from Gaussianity has therefore become a central objective of modern cosmology \cite{Celoria:2018euj_png_in_data,Coulton:2024vot_png_vs_data}.

PnG provide a unique window into the physics of the very early Universe. In particular, they offer a powerful discriminator between competing inflationary scenarios and, more generally, between different mechanisms for generating primordial perturbations \cite{Martin:2013tda_encyclo}. PnG typically affect higher-order correlations beyond the two-point function, which are sensitive to field interactions (cosmological collider) \cite{Arkani-Hamed:2015bza_review_cosmo_collider,Sohn:2024xzd_cosmo_collider_in_planck}, non-trivial initial states \cite{Ansari:2024pgq_png_non_bunch_davies}, and departures from slow-roll evolution \cite{Martin:2012pe_png_from_USR}. As a consequence PnG are an essential probe of the very high-energy physics at play during inflation that is inaccessible through two-point statistics alone.

Importantly, the imprints of PnG propagates across a wide range of cosmological observables and physical scales. Their effects have been extensively studied in the CMB through the bispectrum and trispectrum \cite{Planck:2019kim_PNG_constr,Sohn:2024xzd_cosmo_collider_in_planck}, but they also leave characteristic signatures in the large-scale structures of the Universe \cite{Desjacques:2010nn_png_in_LSS,Assassi:2015jqa_png_in_LSS}.
Beyond the aforemetioned correlation functions, PnG directly modify the distribution of fluctuations, especially in the tail -- associated to very strong fluctuations. This affects the statistical abundances of collapsed objects and underdense regions alike, affecting the abundance and morphology of halos \cite{Maggiore:2009rx_HMF_III,DAmico:2010ywu_non_perturbative}, clusters, and cosmic voids \cite{Kamionkowski:2008sr_before_Damico,Lam:2009nd_before_DAmico,DAmico:2010dwy_void_png,DAmico:2010ywu_non_perturbative}. Understanding these signatures in a unified framework is therefore crucial for fully exploiting the constraining power of upcoming surveys.

At the same time, PnG pose a significant theoretical challenge. Their complete description goes beyond standard perturbative approaches and motivates the development of more sophisticated analytical tools. Depending on the regime and the physical problem under consideration, a variety of techniques have been employed, including the $\delta N$ formalism \cite{Sasaki:1995aw_seminal_deltaN,Pi:2022ysn_deltaN_pdf,Tada:2016pmk_deltaN_bispectrum,Gow:2022jfb_nonpert_PBH_1pt_func}, stochastic inflation \cite{Vennin:2015hra_deltaN_and_stochastic,Cruces:2022imf_review_stochastic_inflation}, or path-integral methods \cite{Maggiore:2009rv_HMF_I,DAmico:2010ywu_non_perturbative}. Although these approaches have led to major advances in our understanding of the generation and evolution of non-Gaussian fluctuations, obtaining robust predictions for strongly non-Gaussian tails of the density distribution remains difficult, especially when one aims to connect primordial physics to late-time observables.

In this work we use the excursion set formalism, the standard analytical framework for describing the formation and abundance of both halos and voids. Owing to its conceptual simplicity and flexibility, it has become an essential ingredient of structure formation theory \cite{Zentner:2006vw_thesis_excursionset}. At the same time, several caveats are well known, including issues related to non-Markovian corrections \cite{Maggiore:2009rv_HMF_I,Maggiore:2009rx_HMF_III}, barrier ambiguities \cite{Maggiore:2009rw_HMF_II,DeSimone:2010mu_moving_bar,Auclair:2026tfy_blachier_movinbgar}, filtering dependencies \cite{Zentner:2006vw_thesis_excursionset,Schneider:2013ria_testing_laceycole}. Despite these limitations, excursion set theory often provides an excellent starting point to construct physically motivated and computationally tractable predictions for structure abundances.

One of the central difficulties of the excursion set approach concerns the incorporation of non-Gaussian statistics. While perturbative non-Gaussianities, notably of the local $f\subsc{NL}$ type, have been investigated within this formalism (see Refs. \cite{DAmico:2010dwy_void_png,Maggiore:2009hp_application} and references therein), it remains highly non-trivial to extend these treatments to fully non-perturbative probability tails \cite{DAmico:2010ywu_non_perturbative}, which are quite common in inflationary models \cite{Cruces:2025typ_deltaN_pdf,Ezquiaga:2022qpw_elgordo_fourier_PNG,Coulton:2024vot_png_vs_data,Gow:2022jfb_nonpert_PBH_1pt_func}. In particular, rare-event statistics are precisely the regime where perturbative expansions tend to break down, and where the sensitivity to PnG becomes the strongest. Capturing the full structure of these tails therefore requires methods that go beyond standard moment expansion.

The purpose of the present work is to explore strongly non-Gaussian tails by the introduction of large deviation techniques. Large deviation theory \cite{dembo2009_zeitouni_book_LDP} provides a natural framework for describing exponentially suppressed fluctuations and rare events \cite{Touchette_2009_review,Burenev_2025_touchette_recent}. In this sense, it offers a promising bridge between primordial non-Gaussianities and excursion-set-inspired approaches to structure formation. Although large deviation methods have already appeared in various areas of cosmology \cite{Cohen:2022clv_LDP_for_inflation,Uhlemann:2015npz_LDP_cosmicfield,Uhlemann:2017tex_nongauss_spheres,Bernardeau:2015khs_LDP_for_LSS}, they have not yet been systematically developed within the context of cluster and void statistics.

More specifically, the goal of this article is to establish a proof of concept for the use of the large deviation principle (LDP) in the determination of the structure abundances. We will show how it can help break some of the theoretical challenges evoked in this introduction. In future work, we aim to extend the framework developed here in two complementary directions: on the one hand, by connecting the large deviation description to specific inflationary models and their associated primordial statistics; on the other hand, by linking these theoretical predictions to numerical simulations and to observational probes of halos and voids in large-scale structure data.

A further motivation for the present work arises from the study of cosmic voids. Compared to dark matter halos and clusters, cosmic voids have historically received less attention, despite their clear growing importance as cosmological probes \cite{Pisani:2015jha_voids_for_DE,Voivodic:2016kog_void_in_MG}, see especially the reviews \cite{Pisani:2019cvo_whitepaper,Contarini:2026yfv_pisani_recent_review}. Void statistics are particularly sensitive to both gravitational dynamics and the properties of the initial conditions, making them promising observables for constraining PnG \cite{Uhlemann:2017tex_nongauss_spheres}. For this, we require accurate theoretical predictions for the void size function \cite{Jennings:2013nsa_Vdn_model,Ronconi:2019xex_improving_vdn,Verza:2024rbm_movingbar,Verza:2024ilg_voidtheory_fits_sims} and for the abundance of rare underdense regions across a broad range of scales.

Significant progress has nevertheless been achieved in the modelling of voids over the past decades. Early foundational work by \cite{Sheth:2003py_seminal} establishes the excursion-set description of void formation and highlights the importance of the void-in-cloud process. More recently, refined models such as the volume-conserving Vdn prescription \cite{Jennings:2013nsa_Vdn_model,Verza:2019tvg_pisani_for_Vdn}, as well as moving-barrier implementations \cite{Ronconi:2019xex_improving_vdn,Verza:2024rbm_movingbar}, have improved the agreement between analytical predictions and numerical simulations \cite{Ronconi:2019xex_improving_vdn,Hamaus:2014fma_VSF_from_sims,Verza:2024ilg_voidtheory_fits_sims,Contarini:2026yfv_pisani_recent_review}. These developments have demonstrated the richness of void phenomenology and the necessity to incorporate increasingly realistic dynamical ingredients into theoretical descriptions.

The presnt article is organized as follows. In section \ref{sec:excursionset} we briefly review the main ingredients of the excursion set formalism and the difficulties of its implementation for non-Gaussian distributions. We also shed light on the implicit assumptions behind some standard results. We introduce the large deviation principle in section \ref{sec:LDP} and discuss some simple but relevant applications in section \ref{sec:appli}, considering the abundance of dark matter halos. Section \ref{sec:voids} is dedicated to voids, where we obtain a new prediction for the void size function. In section \ref{sec:opening} we discuss our results and link the present work to other relevant fields, before concluding. Although they constitute a significant part of our results, mathematical proofs are relegated to Appendices~\ref{appx:homo}--\ref{appx:random-walks}.

\paragraph{Notations.} 

The probability of some event $A$ is denoted $\Pro(A)$. For clarity, random variables and random fields are denoted with uppercase letters and their values with lowercase, e.g., $\Pro(\Delta=\delta)$. The probability density of a random variable $Y$ at $y$ is denoted $p_Y(y)$. We use Fourier convention 
\begin{equation}
    f(x) = \int \frac{\dd^3 k}{(2\pi)^3} e^{ikx} f_k\mperiod
\end{equation}
Given some very large real space volume $V$, we define the volume of a Fourier cell $\vbar^{-1} \equiv (V/(2\pi)^3)^{-1}$. We then interchangeably use $\dps \sum_k$ and $\dps \vbar\!\! \int\! \dd^3 k$. Finally, we introduce the notation $\Ralft$ to refer to ``half" of $\R^3$, i.e., 
\begin{equation}
    \Ralft=\{(k_1,k_2,k_3)\in\R^3\ |\ k_3 > 0 \ \text{or}\ (k_3=0\ \text{and}\ k_2>0) \ \text{or}\ (k_3=k_2=0 \ \text{and}\ k_1 \gs 0)\} 
\end{equation}
and $\Kalf(R)$ the subset of $\Ralft$ with vectors of norm $\abs{k} \ls 1/R$. More details are provided in Appendix \ref{appx:homo}.

\section{Cosmological random walks}
\label{sec:excursionset}

\subsection{Excursion set formalism}
\label{sub:excursionset}

The original Press--Schechter \cite{1974_press_schechter_seminal} approach provides a remarkably simple solution to the problem of predicting  the abundance of dark matter halos, but it suffers from several conceptual shortcomings. Most notably, the derivation relies on an ad hoc factor of $2$ to account for underdense regions and does not properly track the hierarchical nature of gravitational collapse (the so-called ``cloud-in-cloud" effect). Excursion set theory resolves these issues by reformulating the problem as a stochastic process in which the smoothed density contrast evolves with filtering scale. Let us first briefly review the core ingredients of this formalism. 

We fix some redshift $z$ that we leave as an implicit argument to all quantities we now introduce. If $\delta(x) = (\rho(x) - \bar{\rho})/\bar{\rho}$ is the total matter density contrast in the comoving gauge\footnote{Although all cosmological gauges become equivalent on sub-Hubble scales, rigorously $\delta$ should be understood in the comoving gauge. In this gauge the matter is at rest \cite{Durrer:2020fza_book}, so over- and under-densities are orrectly defined with respect to the matter flow. See also Section 3.7 of Ref. \cite{Auclair:2020csm_comoving_gauge} for further motivations.}, we define the contrast smoothed on a scale $R$ around $x$ as
\begin{equation}
\label{eq:window-realspace}
\delta_x(R)=\int \dd^3y\, W(\abs{x-y}, R)\delta(y)  \mcomma
\end{equation}
where $W$ is a window function (or filter) that we will specify below. In Fourier space,
\begin{equation}
\label{eq:window-fourierspace}
\delta_x(R)= \int \frac{\dd^3 k}{(2\pi)^3} e^{ikx}  \widehat{W}(k,R) \delta_k \mperiod
\end{equation}
It is customary to introduce the variance of the smoothed density field,
\begin{equation}
\label{eq:powerspectrum}
\avg{\Delta^2(R)}
= \int
 \frac{\dd ^3k}{(2\pi)^3}
P\subsc{lin}(k,z)\,|\widehat{W}(k,R)|^2\mcomma
\end{equation}
where $\Delta(R)$ denotes a random realization of the smoothed field $\delta(R)$ (we now drop the label $x$). Here, $P\subsc{lin}(k,z)$ is the matter power spectrum of the linear theory at redshift $z$. 
We also introduce the typical mass $M$ inside a region of size $R$,\footnote{$V(R)$ is unrelated to the volume $V$ introduced in the notations section to discretize Fourier space. Disambiguation should be clear from context.}
\begin{equation}
    \label{eq:R-vs-M}
    M = \bar{\rho}_{0,m} V(R) =\frac{4}{3}\pi \bar{\rho}_{0,m} (cR)^3 \mcomma
\end{equation}
where $\bar{\rho}_{0,m}$ is the mean matter density today, and the constant $c$ notoriously depends on the filter $W$. Anticipating our choice of Fourier top-hat for the filter, in this article we use the Lacey-Cole convention $\frac{4}{3} \pi c^3 = 6\pi^2$, i.e., $c\simeq 2.42$ \cite{1993MNRAS.262..627L_laceycole_constant}. This value is close to the one obtained in N-body simulations, $c\simeq 2.5 -2.7$, making it a motivated theoretical choice \cite{Schneider:2013ria_testing_laceycole}.\footnote{As thoroughly explained in Ref. \cite{Maggiore:2009rv_HMF_I}, the Fourier top-hat filter in principle leads to an ill-defined relationship between $M$ and $R$. Possible ways out are to set $c=1$ like for the real space top-hat, to the detriment of consistency, or to use the Lacey-Cole convention as we do here.} 

\begin{figure}
    \centering
    \includegraphics[width=0.48\linewidth]{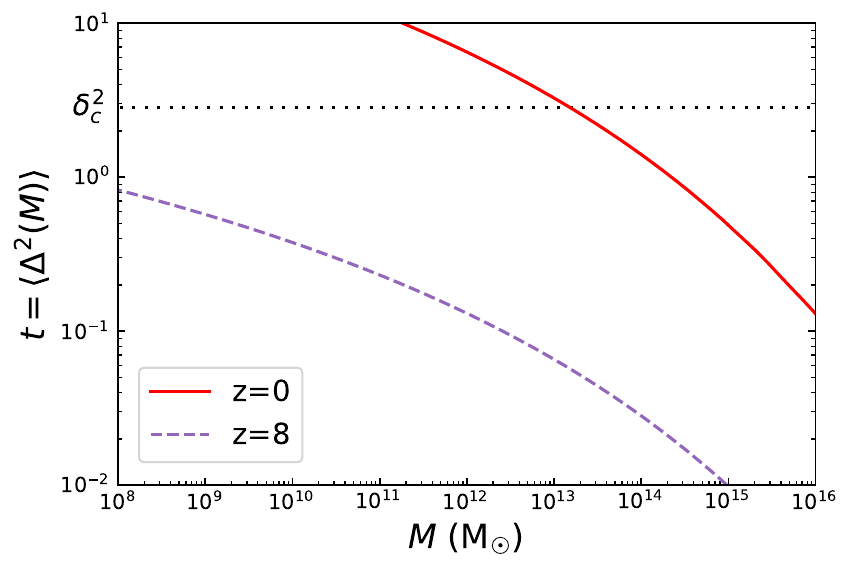}
    \includegraphics[width=0.48\linewidth]{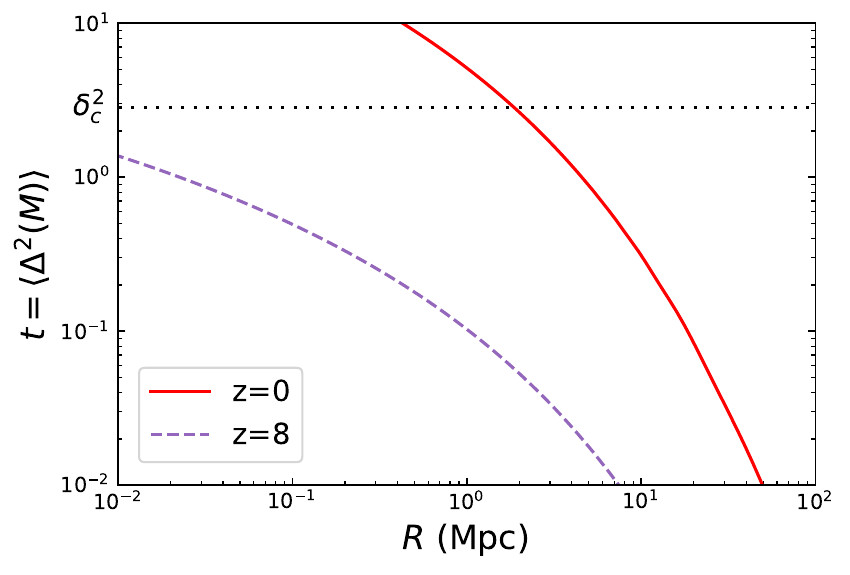}

    \caption{Evolution of the linear theory variance \eqref{eq:powerspectrum} with mass scale $M$ (left) and smoothing scale $R$ (right). The computation is done for a Fourier top-hat filter \eqref{eq:def-tophat} using \texttt{CAMB} \cite{Lewis:1999bs_CAMB} with $\Lambda$CDM parameters \cite{Planck:2018vyg_cosmo_parameters}. $M$ and $R$ are linked through \Eq\eqref{eq:R-vs-M}. The validity of the large deviation approach, which we will introduce in Section \ref{sec:LDP}, corresponds to scales for which the associated variance falls below $\delta_c^2 \simeq 2.8$ for halos (or $\delta_v^2 \simeq 7.4$ for voids). At $z=0$, this threshold corresponds to $R\simeq \text{few}\ \mathrm{Mpc}$ and $M\simeq \text{few}\ 10^{13}\solarmass$; at $z=8$,  the threshold is below $R\simeq 10^{-3} \mathrm{Mpc}$ and $M \simeq 10^5\solarmass$.}
    \label{fig:variance-to-R}
\end{figure}

In the excursion set formalism, one considers random walks of the smoothed density contrast $\Delta(R)$  at fixed redshift $z$, as the smoothing scale $R$ starts at $R=\infty$ ($\Delta(R=\infty)=0$) then decreases. Rather than $R$ itself, the stochastic ``time'' $t$ of excursion set theory is set to the variance of the smoothed field,
\begin{equation}
t = S(R) \equiv \avg{\Delta^2(R)} \mcomma
\end{equation}
the correspondence between $S(R)$ and $R$ being monotonic.  Since smaller smoothing scales probe stronger fluctuations, $t$ increases as $R$ decreases \cite{Bond:1990iw_seminal}. This well-known relationship, shown in Figure \ref{fig:variance-to-R}, can be computed with approximate analytical expressions \cite{Eisenstein:1997jh_power_spectrum} or with a software. In our analysis we use the \texttt{CAMB} package \cite{Lewis:1999bs_CAMB}.

In this framework, halo formation (see Section \ref{sec:voids} for cosmic voids) is determined by the first ``time" a trajectory $\De(t)$ crosses a collapse barrier $b$. This barrier is usually taken to be the critical overdensity $\de_c=1.686$ of spherical collapse in the linear theory \cite{peebles:1971_book}. For a given $b>0$ (respectively $b<0$), we denote $T_b = \inf_{t>0}\{t| \De(t)\gs b\}$ (respectively $T_b = \inf_{t>0}\{t| \De(t)\ls b\}$) the first-passage time of the random walk $\De$ at $b$ (recall $\De(t=0)=0$). The first-passage time (FPT) distribution $f\fpt(t) \dd t = p_{T_b}(t)\dd t$ gives the fraction of trajectories crossing the barrier for the first time between $t$ and $t+\dd t$. Although many theoretical improvements can be considered (e.g. moving barriers \cite{Maggiore:2009rw_HMF_II,DeSimone:2010mu_moving_bar,Verza:2024rbm_movingbar}, non-spherical collapses \cite{Sheth:2001dp_tormen_ellips},...), in the standard approach this quantity is directly related to the comoving halo mass function  by \cite{1993_padmanabhan_book,Zentner:2006vw_thesis_excursionset} 
\begin{equation}
\label{eq:HMF}
\frac{\dd n(z)}{\dd \ln M} = \frac{\bar{\rho}_{0,m}}{M} f\fpt(t(z))\frac{1}{3} \left|\frac{\dd t(z)}{\dd \ln R}\right| \mcomma
\end{equation}
with $R$ obtained from $M$ via \Eq \eqref{eq:R-vs-M}.

\subsection{First-passage time beyond the Gaussian assumption}
\label{sub:beyond-gauss}

A particularly important choice of window function $W$ that we use in the rest of this article is the sharp Fourier-space top-hat filter,
\begin{equation}
\label{eq:def-tophat}
\widehat{W}(k,R)=\Theta(1-\abs{k}R) \mcomma
\end{equation}
where $\Theta$ is the Heaviside step function. It is often read that such a filter greatly simplifies computations, as new increments to the random walk between $t$ and $t+\dd t$ now become independent, i.e. the walk becomes Markovian \cite{Maggiore:2009rv_HMF_I,Zentner:2006vw_thesis_excursionset}. One must however use this statement carefully. Given statistical homogeneity of cosmological perturbations, the independence of modes $k\neq p$ can only be straightforwardly guaranteed if the field distribution is assumed to be jointly Gaussian (see Appendix \ref{appx:homo}). A core reason why the excursion set formalism is able to provide analytical results in this case is that, either from \Eq \eqref{eq:window-realspace} or \Eq \eqref{eq:window-fourierspace}, $\De(R)$ is known to be Gaussian as a sum of joint Gaussians. Once this Gaussian profile is manifest, one only needs to compute its variance, which is simply given by \Eq\eqref{eq:powerspectrum}.

As mentioned in the introduction, we however expect from inflationary models that large fluctuations of the local density field $\De_k$ do not follow a Gaussian distribution. This severely complicates the issue of determining the distribution of the smoothed field $\De(R)$, as {\it (i)} the independence of the $\De_k$'s is usually lost, even with a Fourier top-hat filter; {\it (ii)} shall independence still remain, a (weighted) sum \eqref{eq:window-fourierspace} of random variables does not abide by any particular property and its distribution may be extremely difficult to compute. 

A question then comes naturally: which results remain when Gaussianity is lost? Let us for this section conserve -- now as an independent assumption -- the Markovianity (i.e. independence of modes) of the random walk. We postpone the question whether this assumption is physical or not to Section \ref{sub:LDP-indep-fourier}.

Most importantly, in this case we can prove an exact connection between the distributions of $T_b(t)$ and $\De(t)$ that does {\it not} rely on Gaussianity. Providing that the random walk {\it (i)} is Markovian (i.e. Fourier modes are independent); {\it (ii)} is symmetric (i.e., $\De(t)$ and $\De(-t)$ have the same distribution); {\it (iii)} satisfies a natural property that $\De(t)$ and $\sqrt{t}\De(1)$ have the same distribution; then $T_b$ has the same distribution as $b^2/\De^2(t=1)$. In other words, the FPT distribution $f\fpt(t)$ is given by

\begin{equation}
\label{eq:1st-passage-time-vs-walk}
    f\fpt(t) \equiv p_{T_b}(t) = \frac{b}{t^{3/2}} p_{\De(1)}\left(\frac{b}{\sqrt{t}}\right) \mperiod
\end{equation}
We defer the proof of this equality to Appendix \ref{appx:sub:basics}. Note that here we are considering random walks with pure noise and no deterministic drift. \Eq \eqref{eq:1st-passage-time-vs-walk} is verified by generic Lévy processes and not specific to the Wiener process, again meaning that no underlying Gaussian distribution needs to be assumed -- but Markovianity does. Combining the above equation with \Eq\eqref{eq:HMF} sets a clear roadmap: if one derives the distribution of the smoothed field $\De(R)$ (e.g. from the pointwise Fourier field $\De_k$ via \eqref{eq:window-fourierspace}), one radily obtains a prediction for the non-Gaussian halo mass function.

\subsection{Connecting Fourier space to real space distributions}
\label{sub:fourier-to-real}

We conclude this section by observing that the distribution of $\De(R)$ can be established analytically from that of the variables $\De_k$, providing the latter are independent. However, because the exact expression is cumbersome to calculate in practice, this reinforces the motivation to use tools from the theory of large deviations.

Let us first provide a more precise meaning of the {\it independence between Gaussian Fourier modes}. We write the smoothed density field around $x$ using a Fourier top-hat filter,
\begin{align}
    \De(R) &= \int_{K(R)} \frac{\dd^3k}{(2\pi)^3} e^{ikx} \De_k& K(R) = \{k \ | \ \abs{k} \ls R^{-1}\}\mcomma \\
    &= 2 {\rm Re} \int_{\Kalf(R)}\frac{\dd^3k}{(2\pi)^3} e^{ikx} \De_k & \Kalf(R) = K(R) \cap \Ralft\mperiod
    \label{e:DeR}
\end{align}
The second equality is a  simple consequence of the reality of $\De$ in real space, hence $\De_{-k}^*=\De_k$ and the integral can be performed only on ``half" of the modes satisfying $\abs{k} \ls R$. This humble observation has an importance consequence: the set of variables $\Delta_k,~k\in  \Ralft$ are mutually independent, i.e., their joint probability is obtained as the product of their individual distributions (see Appendix \ref{appx:homo} for interesting technicalities). Again, what if this independence extends beyond the Gaussian regime? Using it we can connect the characteristic function $\Phi$ of $\Delta(R)$ and the one of the $\De_k$'s,
\begin{align}
\label{eq:link-Phi-Phiks}
    \Phi_{\De(R)}(y) \equiv \avg{e^{i y \De(R)}} = \prod_{k\in \Kalf(R)} \Phi_{\De_k}\left(\frac{2}{(2\pi)^3\vbar} y e^{-ikx}\right) \mcomma
\end{align}
where for a variable $X$ and $z\in \C$, the characteristic function is defined as $\Phi_X(z) \equiv \avg{\exp(i{\rm Re}(z^* X))}$. We have made use of the equality 
\begin{equation}
    \exp \int \dd^3k\, c_k = \exp \sum_k c_k /\vbar = \prod_k e^{c_k/\vbar} \mperiod 
\end{equation}

We further recall that statistical homogeneity of our Universe implies that {\it (i)} the phase of $\Delta_k$ is always uniformly distributed; {\it (ii)} its modulus and its phase are always independent random variables.\footnote{Note that this is not the case of its real and imaginary parts. In virtue of Herschel-Maxwell's theorem \cite{mukherjee2017proofherschelmaxwelltheoremusing}, independence of $\Re(\De_k)$ and $\Im(\De_k)$, together with uniformly distributed phases would enforce $\De_k$ to be Gaussian.} Contrary to mode independence, we emphasize that both of these statements are completely general and do not rely on Gaussianity (see again Appendix \ref{appx:homo}). Writing $\De_k = \abs{\De_k} e^{i\Th_k}$ we have $p_{\De_k}(u e^{i\th}) = p_{\abs{\De_k}}(u) / (2\pi u)$, which implies 
\begin{equation}
\label{eq:hankel-transform}
    \Phi_{\Delta_k}(z) \equiv \avg{e^{i{\rm Re}(z^* \Delta_k)}} = \int_0^\infty p_{\abs{\De_k}}(u) J_0(u \left| z \right|)\dd u \mperiod
\end{equation}
Here, $J_0(x) = \frac{1}{2\pi}\int_0^{2\pi} e^{i x \cos\theta}\dd \theta$ is the Bessel function of the first kind (\Eq\eqref{eq:hankel-transform} is close to a Hankel transform of order $0$). Notice that, thanks to the rotational symmetry of $p_{\Delta_k}$, $\Phi_{\De_k}(z)$ depends only on $|z|$. Inserting this in \Eq\eqref{eq:link-Phi-Phiks} provides an exact equation for the distribution of $\De(R)$, given $p_{\abs{\De_k}}$, 
\begin{align}
    p_{\Delta(R)}(\delta) &= \frac{1}{2\pi} \int_{-\infty}^{\infty}\dd y\, e^{-iy\delta}\Phi_{\Delta(R)}
(y) \\
\label{eq:density-from-Bessel}
&=\frac{1}{2\pi} \int_{-\infty}^{\infty}\dd y \exp\left[-i y\delta + 2\pi\vbar \int_0^{1/R} \dd k k^2 \ln\left(\int_0^\infty p_{\abs{\De_k}}(u) J_0\left(\frac{2 y u}{(2\pi)^3 \vbar}\right)\dd u\right)\right] \mperiod
\end{align}
This expression cannot be simplified in general. Let us first estimate the integral over $y$ by a saddle-point approximation, valid for $y \ll \de^{-1}$, near to $y\simeq0$. Expanding $J_0(2x) \simeq 1- x^2$, this leads to
\begin{align}
p_{\Delta(R)}(\delta)
&\simeq\frac{1}{2\pi} \int_{-\infty}^{\infty}\dd y \exp\Big[-i y \delta - \frac{y^2}{2} \underbrace{\frac{2}{(2\pi)^5 \vbar} \int_0^{1/R} \dd k k^2\avg{\abs{\De_k}^2} }_A\Big]  \\
\label{eq:everything-is-gauss}
&= \frac{1}{\sqrt{2\pi A}}e^{- \delta^2/(2A)}\mperiod
\end{align}
This would lead to the conclusion that $\De(R)$ is  Gaussian. As we aim at exploring non-Gaussian distributions, this approximation is clearly not good enough. 

We must seek for another way to link the $p_{\abs{\De_k}}$'s to $p_{\De(R)}$. As stated in Section \ref{sub:beyond-gauss}, here the issue remains that evaluating the distribution of a non-trivial combination of random variables, even independent, is a difficult problem.  In the next section, we explain how the large deviation principle allows us to circumvent this problem and recast it in a completely different, much simpler framework.

\section{Large Deviation Principle (LDP) for random fields}

\label{sec:LDP}

We now release the assumption of independence between Fourier modes that was made earlier, and will restore it in Section \ref{sub:LDP-indep-fourier}.

\subsection{Modelling rare fluctuations}
\label{sub:LDP-intro}

The large deviation principle (LDP) is a powerful formalism for the study of rare events, built to capture the essential features of the tails of probability distributions. It is therefore well-suited to model strong density fluctuations in our Universe that may lead to the formation of very large clusters or voids, the typical scales of which are illustrated in \fig\ref{fig:variance-to-R}. As these  rare fluctuations are often most sensitive to non-Gaussian initial conditions coming from inflation \cite{Ezquiaga:2022qpw_elgordo_fourier_PNG}, the LDP may allow us to more accurately connect initial conditions to the observation of rare cosmic structures. Through a few key examples, we attempt to show how this connection is made possible and what sort of theoretical challenges can be addressed with this tool. 

In the main text, we introduce minimal ingredients required for this article to be self-contained. More general definitions are provided in Appendix \ref{appx:LDP}, but the theory of large deviations is of course much richer. The reader interested in an exhaustive, formal treatment will find much depth in the acknowledged book of Dembo and Zeitouni \cite{dembo2009_zeitouni_book_LDP}. A more accessible, application-oriented introduction can be found in the reviews of Touchette {\it et al.} \cite{Touchette_2009_review,Burenev_2025_touchette_recent}.  While the mathematical concepts introduced here are taken from standard literature, their application within our framework is new.  The reader mainly interested in cosmological observables may directly skip to Sections \ref{sec:appli} and \ref{sec:voids} where we present the applications to cluster and void statistics. 

\subsubsection{Mathematical introduction}
\label{ssub:LDP-intro}
We say that the Fourier density contrast $\De_k$ with values in $\C$ {\it satisfies a LDP} if its probability density depends on a parameter $\eps_k>0$ such that
\begin{equation}
\label{eq:def-rate-func-single-k}
    p_{\De_k}(z_k)\underset{\eps_k \to 0}{\asymp} \exp(- I_k(z_k)/\eps_k) \mcomma
\end{equation}
where $\asymp$ denotes an equivalence up to sub-exponential factors, i.e. $p(z) = \exp(-I(z)/\eps + o(1/\eps))$, and $I_{\De_k}$ (abbreviated $I_k$) is called the {\it rate function} of $\De_k$. Since our goal is to study large fluctuations, i.e. $\abs{\De_k} \gg \avg{\abs{\De_k}}$, the parameter $\eps_k\to 0$ will be related to the variance of the probability density that tends to zero when $R\to\infty$.  We will further need to consider the {\it joint} distribution of the $\De_k$'s. Let us denote with bold letter $\bm{\De} \equiv \{\De_k\}_k$ the joint set of all $\De_k$'s for $k\in \Kalf(R)$, taking values $\bm{z} = \{z_k\}_k \in L^2(\Kalf(R))$. The joint rate function is similarly defined such that
\begin{equation}
\label{eq:def-joint-rate}
    p_\bde(\bm{z}) = p_{\De_{k_1},\De_{k_2},\dots} (z_1,z_2,\dots)\underset{\eps \to 0}{\asymp} \exp(- I\subsc{joint}(\bm{z})/\eps) \mcomma
\end{equation}
with $\eps = \sup_k \eps_k = \norm{\bm{\eps}}_\infty$.

Our objective is to determine the (non-Gaussian) distribution of $\De(R)$, the main unknown affecting the halo mass function. This goal can now be achieved by using a tool called the {\it contraction principle}. Let us first describe it with a toy example: what is the connection between the rate functions of $\De_k$ and $\abs{\De_k}$? Because there exists a map $\abs{\,\cdot\,}:z\mapsto \abs{z}$ linking both variables, the contraction principle states that the rate function $I_{\abs{\De}}$ of the latter can be obtained from (we drop the index $k$ for a few lines)
\begin{equation}
    I_{\abs{\De}}(u) = \inf\, \{ I_{\De}(z) \ , \ z\in \C \ \text{such that}\ \abs{z} = u\}\mperiod
\end{equation}
Recall that from mere statistical homogeneity,  the distribution of $\De_k$ depends only on its modulus, since its phase must be uniformly distributed (see Appendix \ref{appx:homo}). Thus, $I_\De(z)$ is a function of $\abs{z}$ only, so that in this specific example we find
\begin{equation}
\label{eq:ratefunction-rot-invariance}
    I_{\abs{\De}}(u) = \inf_{z\in \C, \abs{z}= u} I_{\De}(\abs{z}) = I_{\De}(u)\mperiod
\end{equation}
We conclude that $\abs{\De_k}$ and $\De_{k}$ share the same rate function; we will use this fact throughout.

Pushing forward rate functions through maps is the essence of the contraction principle. This has even more interesting applications. Because $\De(R)$ is a function of the $\De_k$'s,
\begin{equation}
\label{eq:De(R)-func-of-De(k)}
    \De(R) = F(\{\De_k\}_k) = 2 {\rm Re} \int_{\Kalf(R)}\dd^3k\, e^{ikx} \De_k\mcomma
\end{equation}
if the $\De_k$'s satisfy a LDP it is guaranteed that $\De(R)$ must satisfy one as well. That is, there exists a parameter $\eps(R)$ (we will specify it in Section \ref{sub:expo-tails}) such that
\begin{equation}
    \label{eq:LDP-on-De(R)}
    p_{\De(R)}(\delta) \underset{\eps(R) \ll \delta}{\asymp} \exp(-I_R(\delta) / \eps(R)) \mperiod
\end{equation}
The contraction principle relates the rate functions via
\begin{equation}
\label{eq:rate-functions-link}
    I_{\De(R)}(\de) \equiv I_R(\delta) = \inf_{\bm{z}\in L^2(\Kalf(R))}\left\{I\subsc{joint}(\abs{\bm{z}}) \, \mid  F(\bm{z}) = F(\{z_k\}_k) = \delta \right\} \mperiod
\end{equation}
Here, $\abs{\bm{z}} \equiv \{\abs{z}_k\}_k$.
\Eq \eqref{eq:rate-functions-link} is pivotal: it recasts the problem of finding the distribution of $F(\{\De_k\})$ into an optimization problem in a high (infinite)  dimensional space, regardless of how intricate the function $F$ may be. The advantages are twofold. First, even if the sum \eqref{eq:De(R)-func-of-De(k)} of non-Gaussian variables does not satisfy specific properties, the tail of its distribution can nonetheless be estimated. Second, \Eq\eqref{eq:rate-functions-link} does not require independence between $k$ modes (but requires knowledge of the joint distribution). Note that, at this stage, the correlations between modes are unspecified. Thus, it is as well valid if the joint probability were given for the real space field $\De(x)$ instead,
\begin{align}
     I_R(\delta) &= \inf_{\{\eta_x\}_x}\left\{I\supsc{real}\subsc{joint}(\{\eta_x\}_x) \, \mid  G(\{\eta_x\}_x) = \delta \right\} \mcomma  \intertext{where}
      \De_x(R) &= G(\{\De(y)\}_y) \equiv \int \dd^3 y W(\abs{x-y})\De(y) \mperiod
\end{align}

The next step  forward is to compute the joint rate function. In many cases \cite{Burenev_2025_touchette_recent}, it can formally be obtained with a second tool, the {\it Gärtner-Ellis lemma}, which expresses it as a Legendre transform. The Gärtner-Ellis lemma reads 
\begin{equation}
\label{eq:gartner-ellis}
    I\subsc{joint}(\abs{\bm{z}}) = \sup_{\bm{v}\in L^2(\Kalf(R))} \{ \Re(\bm{z}\cdot \bm{v}) - \Lambda(\bm{v})\}
\end{equation}
where 
\begin{equation}
    \label{eq:def-hermitian-product}
    \bm{z}\cdot  \bm{v} \equiv \int_{\Kalf(R)}\dd^3 k z_k v_k^* \mcomma
\end{equation}
and $\Lambda$ is the scaled cumulant generating function \cite{Burenev_2025_touchette_recent},
\begin{equation}
    \Lambda(\bm{v}) = \lim_{\eps\to 0} \eps\ln\avg{\exp( \Re(\bm{v}\cdot \De)/\eps)} \mperiod
\end{equation}
It is worth noticing that for convex rate functions, the Legendre transform is involutive, 
\begin{equation}
    \Lambda(\bm{v}) = \sup_{\bm{z}\in L^2(\Kalf(R))} \{ \Re(\bm{z}\cdot \bm{v}) - I\subsc{joint}(\abs{\bm{z}})\}\qquad \text{if $I\subsc{joint}$ is convex.}
\end{equation}

\subsubsection{Effect of scale correlations}
\label{ssub:correlations}

The final step is to evaluate $\La(\bm{v})$, which is generally the biggest obstacle \cite{Burenev_2025_touchette_recent}. Here, the correlations between different scales $k$ play a predominant role. A first approach consists of expanding $\La(\bm{v})$ into correlators,
\begin{equation}
\label{eq:correl-expansion}
    \avg{\exp( \Re(\bm{v}\cdot \De)/\eps)} = \sum_{n=0}^\infty \frac{1}{2^n\eps^n n!}\int_{\Kalf(R)} \dd^3k_1\cdots \dd^3 k_n \sum_{b\in\{0,1\}^n} \prod_{i=1}^n v_{k_i}^{(1-b_i)}\avg{\prod_{i=1}^n \De_{k_i}^{(b_i)}} \mcomma
\end{equation}
where $z_i^{(b_i)}$ means $z_i$ whenever $b_i=0$ and $z_i^*$ whenever $b_i=1$. Another strategy is to perform an Edgeworth-like expansion \cite{Lam:2009nd_before_DAmico} of the joint probability, without relying on the Gärtner-Ellis lemma. One writes (for simplicity we directly consider the distribution of the  joint moduli $\abs{\bm{\De}} = \{\abs{\De_k}\}_k$) $p_{\abs{\bde}}$ using the probability distributions $p_{\abs{\De_k}}$'s of each individual mode,
\begin{equation}
\label{eq:edgeworth-exp}
    p_{\abs{\bde}}(\abs{\bm{z}}) = \left(\prod_{k\in \Kalf(R)} p_{\abs{\De_k}}(\abs{z_k}) \right) \times\left[1 + \sum_{\bm{m}=\{m_k\}_k} C(\bm{m})\prod_{k\in \Kalf(R)} Q_{k}^{(m_k)}(\abs{z_k})\right] \mperiod
\end{equation}
This expansion consists of adding order-by-order corrections to the joint probability, stemming from correlations, starting from independent Fourier modes. Let us detail our notations. The tuple $\bm{m}$ runs over all sequences of non-negative integers $(m_{k_1},m_{k_2},\dots)\in \N^{\Kalf(R)}$ with only finite numbers of non-zero coefficients (and such that not all of them equal zero, since we have singled out $1$ from the sum in \Eq\eqref{eq:edgeworth-exp}). For each mode $k\in \Kalf(R)$, one builds with the Gram-Schmidt algorithm a family $\{Q_k^{(m_k)}\}_{m_k}$ of polynomials that are orthonormal for $p_{\abs{\De_k}}$. That is, $\forall k, \forall m_k\in \N, \, \deg Q_k^{(m_k)} = m_k$, $Q_k^{(0)}=1$ and
\begin{equation}
    \int_0^\infty \dd u\, p_{\abs{\De_k}}(u)\, Q_k^{(m_k)}(u) Q_k^{(m'_k)}(u) = \delta^K_{m_k m_k'}\mcomma
\end{equation}
where $\de^K$ is the Kronecker delta.\footnote{In applications below we consider $p_{\abs{\De_k}}(u) = q /(\sig_k \Gamma((1+\beta)/q)) \times (u/\sig)^\beta \exp(-(u/\sig)^q)$, see \Eq \eqref{eq:exp-distrib-fourier}. Let us observe that in this case the polynomials $Q_k$ have an explicit form based on generalized Laguerre polynomials $L^{(\alpha)}_n(X)$: $Q_k^{(m_k)}(X) =\sqrt{\Ga(\vartheta+1)} L^{(\vartheta)}_{m_k}(X^q/\sig^q)$ with $\vartheta = -1 + (\beta+1)/q$.} This is very similar to a generic moment expansion in powers of $\avg{\abs{\De}^n}$. However, the advantage of using orthonormal polynomials is that the coefficients $C(\bm{m})$ of the Edgeworth expansion simply read
\begin{equation}
    C(\bm{m}) = \avg{\prod_k Q_k^{(m_k)}(\abs{\De_k})} \mperiod
\end{equation}
Note that this correlation is always of finite order, owing to the definition of $\bm{m}$ having finite number of non-zero components. Finally, taking the logarithm of \Eq\eqref{eq:edgeworth-exp} allows to write the joint rate function as
\begin{equation}
    \label{eq:rate-func-plus-correc}
    I\subsc{joint}(\abs{\bm{z}}) = \sum_k I_k(\abs{z_k}) \quad + \quad \text{corrections from non-independence}\mperiod
\end{equation}

To make use of either expression \eqref{eq:correl-expansion} or \eqref{eq:edgeworth-exp}, the inputs required from inflationary theory are therefore the successive correlators, bispectrum, trispectrum... of the density contrast (or more precisely, the gauge-invariant curvature $\zeta_k$). These are precisely the targets of the cosmological collider program \cite{Arkani-Hamed:2015bza_review_cosmo_collider,Sohn:2024xzd_cosmo_collider_in_planck,Aoki:2024jha_cosmo_collider_ghoshal}. In full generality, these expressions are likely untractable and require to be treated perturbatively. However, let us make some quick estimates using \Eq\eqref{eq:edgeworth-exp} to see that they may indeed generate non-negligible corrections to the rate function. We group tuples $\bm{m}$ by the order $c=\sum_k m_k < \infty$. Based on dimensional grounds, $\deg Q_k^{m_k} = m_k$ leads to $Q^{(m_k)}_k(X) \propto X^{m_k}/\avg{\abs{\De_k}^{m_k}}$. We then write
\begin{equation}
    \prod_{k} Q_{k}^{(m_k)}(u_k) \sim \left(\frac{u}{\avg{\abs{\De}}}\right)^{\sum_k m_k} \sim \frac{u^c}{\avg{\abs{\De}}^c}\mcomma
\end{equation}
and
\begin{equation}
    C(\bm{m})=\avg{\prod_k Q_k^{(m_k)}(\abs{\De_k})} \sim \frac{\avg{\abs{\De}^{\sum_k m_k}}}{\avg{\abs{\De}}^{\sum_k m_k}} \sim \frac{\avg{\abs{\De}^c}}{\avg{\abs{\De}}^c} \sim \mathcal{O}(1) \mcomma
\end{equation}
from where
\begin{equation}
    I\subsc{joint}(u) = \sum_k I_k(u) + \eps\ln\left(\sum_{c\gs 0}  \mathcal{O}(u^c/\eps^c)\right) \mperiod
\end{equation}
For a simple numerical prefactor $1/c!$ inside the $\mathcal{O}(u^c/\eps^c)$, the right-most term becomes $\eps \ln \exp(\mathcal{O}(u/\eps)) = \mathcal{O}(u)$. Consequently, it may give a contribution to $I\subsc{joint}$ that does not vanish in the limit $\eps \to 0$.

Nevertheless, because scale correlations are more model-dependent \cite{Renaux-Petel:2015bja_review_png_after_planck2015}, we leave their more careful investigation to future work. As stated above, the present work is intended as a proof of concept of the LDP. This is why we now come back to the assumption of independent modes that was introduced in Sections \ref{sub:beyond-gauss}--\ref{sub:fourier-to-real}.

\subsection{LDP for independent Fourier modes}
\label{sub:LDP-indep-fourier}

The task of estimating the joint rate function \eqref{eq:def-joint-rate} is considerably simplified if one assumes the variables $\De_k$, $k\in \Kalf(R)$ to be mutually independent. As recalled in Appendix \ref{appx:homo}, if these variables are jointly Gaussian, their independence is guaranteed by statistical homogeneity. Hence, a simple possibility to conserve this property for non-Gaussian variables $\De_k$ is to postulate that they are {\it locally} built from joint Gaussian variables $\De^G_k$,
\begin{equation}
\label{eq:local-png-fourier}
    \forall k,\ \De_k = \mathcal{F}[\De^G_k] \mperiod
\end{equation}
This echoes the definition of ``local" primordial non-Gaussianities where {\it local} is usually understood in real space \cite{Pinol:2021nha_thesis},
\begin{equation}
    \label{eq:local-png-real}
    \forall x,\ \De(x) = \mathcal{F}[\De^G(x)] \mperiod
\end{equation}
Both regimes can be thought as two opposite extreme situations, delimiting a broad spectrum of mechanisms explaining the emergence of PnG from initial Gaussian seeds. The latter is often considered within the $\delta N$-formalism \cite{Gow:2022jfb_nonpert_PBH_1pt_func,Cruces:2025typ_deltaN_pdf,Ezquiaga:2022qpw_elgordo_fourier_PNG,Pi:2022ysn_deltaN_pdf} or in perturbative treatments based on $f\subsc{NL}\supsc{loc}$-expansions \cite{Pinol:2021nha_thesis}, but the former has also been considered as an efficient way to capture the physics of non-Gaussian tails \cite{Ezquiaga:2022qpw_elgordo_fourier_PNG, Tada:2021zzj_vennin_pdf_coarsegrained}. The clear advantage of \Eq \eqref{eq:local-png-fourier} versus \Eq \eqref{eq:local-png-real} is that the latter leads to correlations in both Fourier and real space, while the former is free from Fourier space correlations. Despite being simpler, \Eq \eqref{eq:local-png-fourier} already leads to interesting predictions, but we note that this does not fully solve the problem, as inflationary models tend to provide distributions for $\De(x)$ rather than $\De_k$ \cite{Pi:2022ysn_deltaN_pdf}. In future works we hope to address this issue with the LDP. Note that correlations have also been adressed for underdensities and cosmic voids \cite{Paranjape:2011bz_more_ado_corrsteps,Musso:2012qk_corrsteps}.

We further emphasize that even if the assumption \eqref{eq:local-png-fourier} corresponds to the ``lowest order" in correlations, the LDP remains a non-perturbative treatment of non-Gaussianities. In a somewhat simplified picture, non-Gaussianities can emerge from either {\it (i)} non-trivial couplings between different scales; or {\it (ii)} fluctuations at each individual scale following non-Gaussian distributions (these two causes being intertwined in reality). Contrary to a $f\subsc{NL}$-like expansion, the LDP does not require a development around a Gaussian distribution; hence it is always non-perturbative with respect to point {\it (ii)}.

For this article, we henceforth assume the $\De_k$'s, $k\in\Kalf(R)$ to be independent random variables (e.g., because \Eq \eqref{eq:local-png-fourier} is satisfied), ensuring Markovianity for the excursion set formalism. Their joint rate function \eqref{eq:def-joint-rate} then breaks down into
\begin{equation}
    I\subsc{joint}(\abs{\bm{z}}) = \sum_{k\in \Kalf(R)} I_k(\abs{z_k}) \mcomma
\end{equation}
where the $I_k$ can be derived from the individual marginal distributions of each $\De_k$. The contraction principle \eqref{eq:rate-functions-link} now leads to the distribution of $\De(R)$,
\begin{equation}
\label{eq:full-expr-pdR}
    p_{\De(R)}(\delta) \underset{\eps(R)\to 0}{\asymp} \exp\left(-\frac{1}{\eps(R)} \inf_{\bm{z}}\left\{\vbar\int_{\Kalf(R)} \dd^3 k I_k(\abs{z_k})\ \middle|\  2 {\rm Re}(\bm{z}\cdot\bm{e})=\delta\right\}\right)\mcomma
\end{equation}
where we have denoted $\bm{e}=\{e^{-ikx}\}_k$ and $\bm{e}\cdot\bm{z}$ is defined by \Eq\eqref{eq:def-hermitian-product}. From there, result \eqref{eq:1st-passage-time-vs-walk} can be used owing to Markovianity, and eventually we recover \eqref{eq:HMF}.

In the next section, we construct a key family of probability densities for which we compute \eqref{eq:full-expr-pdR} explicitly.

\section{Abundance of halos}
\label{sec:appli}

\subsection{Exponential tails and new halo mass function}
\label{sub:expo-tails}

A generic prediction of inflationary models \cite{Vennin:2015hra_deltaN_and_stochastic,Ezquiaga:2022qpw_elgordo_fourier_PNG} is that the probability distribution of the gauge-invariant curvature perturbation $\zeta$ features {\it exponential tails}, $p(\zeta)\sim\exp(-\lambda\zeta)$ where $\lambda$ is the decay rate. In comoving gauge (and for adiabatic perturbations), $\zeta_k$ is simply related to the matter density contrast $\delta_k$ \cite{maggiore_GW_vol2}. For that reason, we consider distributions for the random density contrast $\De_k$, of the form
\begin{equation}
\label{eq:exp-distrib-fourier}
    p_{\De_k}(\abs{\de_k})  = A_k(\abs{\de_k})\, e^{-\abs{\delta_k}^q/\sig_k^q}\mcomma
\end{equation}
for some given $q\gs 1$ \footnote{The reason behind this restriction rather than $q>0$ is thoroughly discussed in Appendix \ref{appx:sub:lemma}, where we provide results for both $0<q<1$ and $q\gs 1$. In a nutshell, the results we establish depends on whether $x \mapsto x^q$ is convex or concave.} (let us recall that from statistical homogeneity, $\De_k\in\C$ has a circularly symmetric distribution).  The case $q=2$ corresponds to the Gaussian one, and $q=1$ to an exponential decay. Here, $A_k$ is any power-law function\footnote{\label{foot:powerlaw}In this article the term ``power-law" is used to refer to any function that is subdominant compared to exponentials, and such that the distribution remains correctly normalized. For instance if $A_k(\abs{\de_k}) = \abs{\de_k}^\beta$, $\beta > -1$ is required.} of $\abs{\de_k}$ and $\sig_k = \mathcal{O}(\avg{\abs{\De_k}})$ up to  a numerical factor. We readily see that such a distribution admits a rate function \eqref{eq:def-rate-func-single-k} of the form
\begin{equation}
\label{eq:rate-func-exp}
    I_k(\abs{\de_k}) \equiv -\lim_{\eps_k\to 0}\eps_k \ln p_{\De_k}(\abs{\de_k}) = \abs{\de_k}^q \mcomma
\end{equation}
where we have identified the small parameter of the LDP as
\begin{equation}
    \eps_k = \sig_k^q \mperiod
\end{equation}
Let us make a few comments. First, the regime of large deviations $\eps_k\to 0$ here corresponds to $\De_k \gg  \avg{\abs{\De_k}} \simeq \eps_k^{1/q}$. Second, because the LDP is insensitive to sub-exponential factors, we observe that all choices for the power-law $A_k(\abs{\de_k})$ lead to the same rate function (its contribution vanishes in the limit $\eps_k \to 0$). Although the formula \eqref{eq:rate-func-exp} can be obtained by  a simple limit operation, for completeness we show in Appendix \ref{appx:sub:basics} that the same expression is obtained when using lemma \eqref{eq:gartner-ellis}. 

The goal is now to determine the distribution of the smoothed density contrast, under the assumption of mode independence (see Section \ref{sub:LDP-indep-fourier}).  For the family of probabilities \eqref{eq:exp-distrib-fourier}, the LDP turns this problem into an optimization problem that is exactly solvable. This is one of our main results. We rigorously prove in Appendix \ref{appx:sub:lemma} that for $q\gs 1$ \Eq\eqref{eq:full-expr-pdR} leads to 
\begin{equation}
\label{eq:result-of-LDP}
    p_{\De(R)}(\de) \propto \exp\left(-\frac{\abs{\delta}^q}{\eps(R) (4\pi R^{-3} /3)^q} \frac{2\pi \vbar}{3 R^3}\right) \mperiod
\end{equation}
We reabsorb the various coefficients using the normalization of $p_{\De(R)}$. In addition, because \Eq\eqref{eq:result-of-LDP} is obtained from the LDP, the prefactor in front of the exponential could in principle contain any power-law function of $\delta$.\footnote{See footnote \ref{foot:powerlaw}. For the sake of clarity we will limit our analysis to monomial power-laws.} This leads to 
\begin{equation}
\label{eq:pdt-with-polynom}
    p_{\De(t)}(\delta) = \frac{q\, \gamma}{2\Gamma((1+\al)/q)\sqrt{t}} \left(\frac{\abs{\delta}\gamma}{\sqrt{t}}\right)^\al\exp\left(-\frac{\abs{\delta}^q \gamma^q}{\sqrt{t}^q}\right) \mcomma
\end{equation}
where $\al>-1$ cannot be determined by the LDP, and $\gamma\equiv (\Gamma((3+\al)/q)/\Gamma((1+\al)/q))^{1/2}$. We have switched to the notations of the excursion set formalism introduced in Section \ref{sub:excursionset}, $t \equiv S(R) \equiv \avg{\Delta^2(R)}$.  The value of $\al$ is not impactful for large deviations and we will use this degree of freedom to our advantage when discussing the void size function in Section \ref{sec:voids}.  For $\al=0$ and $q=2$ we recover exactly the Gaussian result.

Note that the cumbersome normalization factor $\gamma$ cannot be omitted. In order to compare apples to apples when discussing the effect of $q\neq 2$ on FPTs, we must ensure that the notion of ``time" is identical for all random walks associated to different values of $q$. This means that the equality $t=\avg{\De^2(R)}$  must hold independently of $q$, which is indeed satisfied for density \eqref{eq:pdt-with-polynom}. One can also argue that to observe the effect of $q\neq 2$ on the tail one should fix the power spectrum to be identical for that of $q=2$, meaning again the variance $\avg{\De^2(R)} = t$ must be the same for all $q$'s.

\begin{figure}
    \centering
    \includegraphics[width = \linewidth]{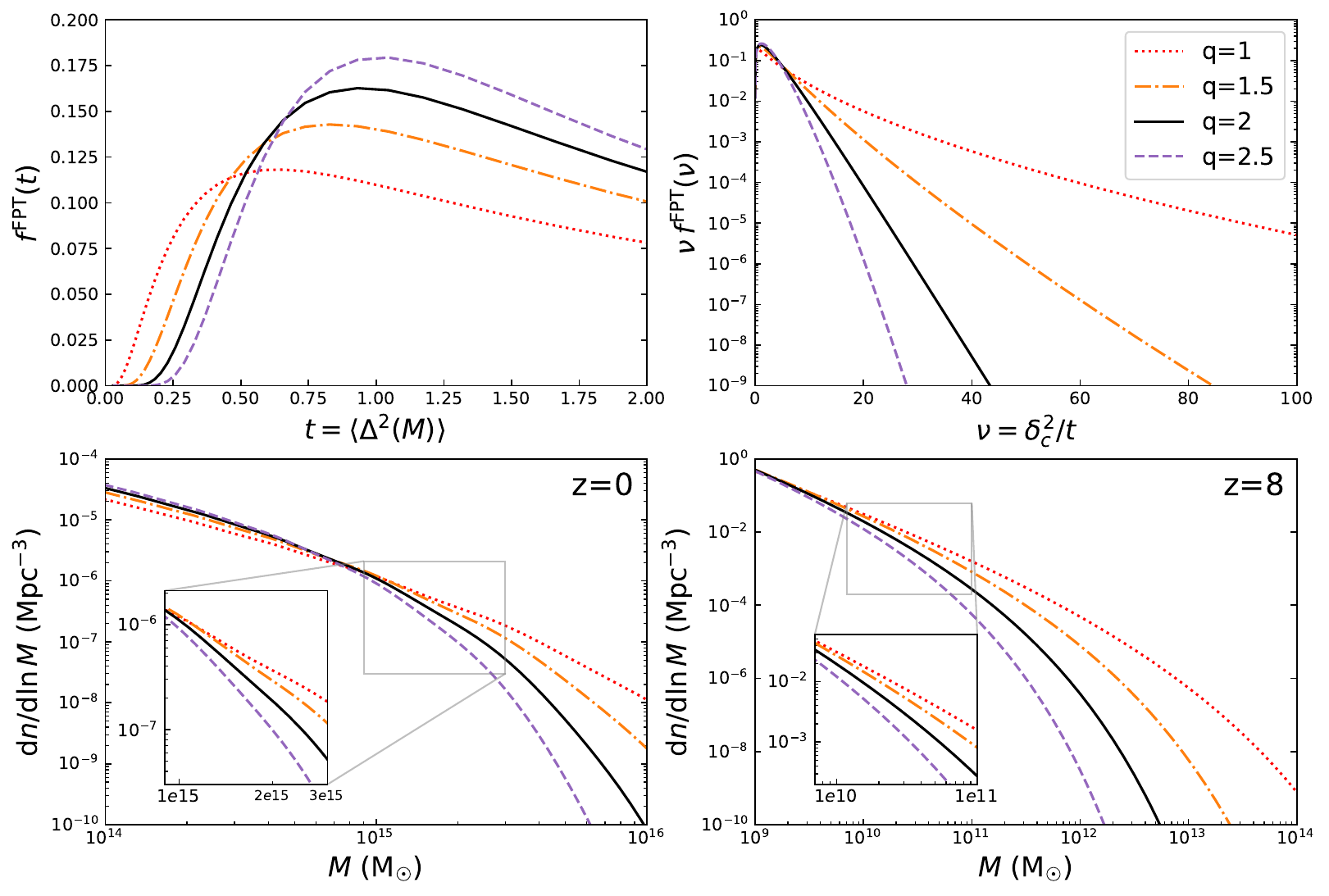}
    \caption{{\it Top left:} Non-gaussian first-passage time distribution \eqref{eq:1st-passage-expo-tail} at $b=\delta_c = 1.686$ for various values of $q$, and $\alpha =0$. {\it Top right:} Same distribution using the customary variable $\nu \equiv \delta_c^2 /t$, i.e. $\nu f\fpt(\nu) = \nu (\dd t/\dd\nu)f\fpt(t)$. {\it Bottom:} The corresponding halo mass functions \eqref{eq:HMF} at redshifts $z=0$ ({\it left}) and $z=8$ ({\it right}). The black line $q=2$ is the standard Gaussian result. Horizontal axis bounds have been chosen to match the regime where the LDP is applicable, $t < \delta_c^2$, see the caption of \fig\ref{fig:variance-to-R}.
    \label{fig:1barrier}}
\end{figure}

As a final step, we obtain the distribution  of the FPT at the collapse barrier $b=\de_c = 1.686$ as a simple combination of \Eqs\eqref{eq:1st-passage-time-vs-walk} and \eqref{eq:pdt-with-polynom},
\begin{equation}
     \label{eq:1st-passage-expo-tail}
      f\fpt(t)=p_{T_{b}}(t)  = \frac{q}{2\Gamma((1+\al)/q)}\frac{b \gamma}{t^{3/2}} \left(\frac{b\gamma}{\sqrt{t}}\right)^\al\exp\left(-\frac{(b\gamma)^q}{\sqrt{t}^q}\right) \mcomma
\end{equation}
Interestingly we can absorb $\ga$ in the height of the barrier. The effect of $q\neq 2$ is thus two-fold: the shape of the tail changes, and the barrier height decreases with $q$ ($\gamma=1/\sqrt{2}$ for $q=2$ and $\al=0$, and it decreases with increasing $q$ for all $\alpha>-1$). 
This distribution is shown in the top panels of \fig \ref{fig:1barrier}. Inserting it in \Eq\eqref{eq:HMF} we infer the halo mass function. We compare our prediction to the standard Gaussian case, $q=2$, in the bottom panels of \fig \ref{fig:1barrier}. 

We observe that the discrepancy with Gaussian fluctuations is stronger in the limit $R\to\infty$ of very large objects. By essence of the LDP, \Eq\eqref{eq:pdt-with-polynom} is most accurate in the large fluctuations regime where the variance is comparably small against the barrier height, $\sqrt{t} = \sqrt{\avg{\De^2(R)}} \ll \de_c$. For $q<2$ the halo mass function is enhanced at very high masses as the probability of large fluctuations is enhanced with respect to the Gaussian case. These large fluctuations increase the fraction of random walks crossing the barrier at lower values of $t$, i.e. larger masses $M$. This effect is hugely significant. For instance, the halo mass function increases by more than $5$ orders of magnitude between $q=2$ and $q=1$ for masses above $10^{13} \solarmass$ at redshift $z=8$.  At a redshift $z=0$, the abundance of very massive halos, $M\sim 10^{16}\solarmass$, is boosted by a factor of a hundred, in agreement with other works not accounting for the cloud-in-cloud problem \cite{Ezquiaga:2022qpw_elgordo_fourier_PNG}, suggesting that this problem is subdominant for very large halos. For $q>2$, the probability of large fluctuations is reduced and we find less very massive halos.

\subsection{Extension to log-normal and other non-symmetric distributions}

In the previous example, the probability distribution \eqref{eq:pdt-with-polynom} is symmetric between over- and under-densities, a condition that is required to apply \Eq\eqref{eq:1st-passage-time-vs-walk}. This approximation is more than satisfying to study halo formation in the linear regime, and in fact is already present in the standard case where a ``Gaussian" -- in particular, symmetric -- distribution is assumed for $\De(R)$. Yet this obviously cannot convey the full picture, since in reality the density fluctuations are bounded from below by $-1$, but not bounded from above. As we now show, it is however not so difficult to extend our results to certain categories of asymmetric distributions. This gives more robustness and more scope to the results presented here.

In some cases where $\De(R)$ does not follow a symmetric distribution, one may still find a change of variable $\phi$ so that $Y(R) = \phi(\De(R))$ is symmetrically distributed. For instance, a common scenario to account for the constraint $\delta>-1$ is to use the log-normal distribution. This distribution has found to be a reasonably good approximation to density fluctuations found in N-body simulations and has also been proposed as the distribution of observed galaxy number counts \cite{Hurtado-Gil:2017dbm_lognormal_for_galaxy}.
 
For this we set $Y(R,\De)=\ln(1+\De(R)) + \sig^2(R)/2$, assuming $Y$ follows a Gaussian distribution with vanishing mean and variance $\avg{Y^2(R)}=\sig^2(R)$. Note the required addition of $\sig^2(R)/2$ since $\avg{\De}=0$ and for a Gaussian variable
\bea
\avg{\exp(Y(R)-\sig^2(R)/2)} &=&\avg{\exp(Y(R))}
\exp(-\sig^2(R)/2) \nonumber \\
&=& \exp(\sig^2(R)/2)
\exp(-\sig^2(R)/2)=1\mperiod
\eea
The distribution of $\De(R)$ is then given by the log-normal distribution
\begin{equation}
    p_{\De(R)}(\delta) = \frac{d y}{d\de}p_{Y(R)}(y(\de))= \frac{1}{\sqrt{2\pi} \sig(R) (1+\delta)}\exp\left(-\frac{[\ln(1+\de)+\sig^2(R)/2]^2}{2\sig^2(R)}\right) \mperiod
\end{equation}
As $Y$ follows a symmetric Gaussian distribution, we can consider a random walk using the variable $Y$ itself. Let us define times $t_y \equiv \sig^2(R)=\avg{Y^2(R)}$ and $t_\delta = \avg{\De^2(R)}$. They are linked by 
\begin{equation}
\label{eq:ty-vs-tdelta}
    1+t_\delta =\exp(t_y) \mperiod
\end{equation}
 As $Y$ is now symmetric, \Eq\eqref{eq:1st-passage-time-vs-walk} applies and the excursion-set formalism provides the distribution of the first-passage time $T_c^y$ of $Y$ at some barrier $c$. Note, however that due to the shift by $\sig^2(R)/2$, the barrier defined by $\de=\de_c$ for $\De$ becomes a moving barrier for $Y$,
\begin{equation}
c = c(\delta_c, t_y) = \ln(1+\de_c) +t_y/2 \mperiod
\end{equation}

\begin{figure}
    \centering
    \includegraphics[width=\linewidth]{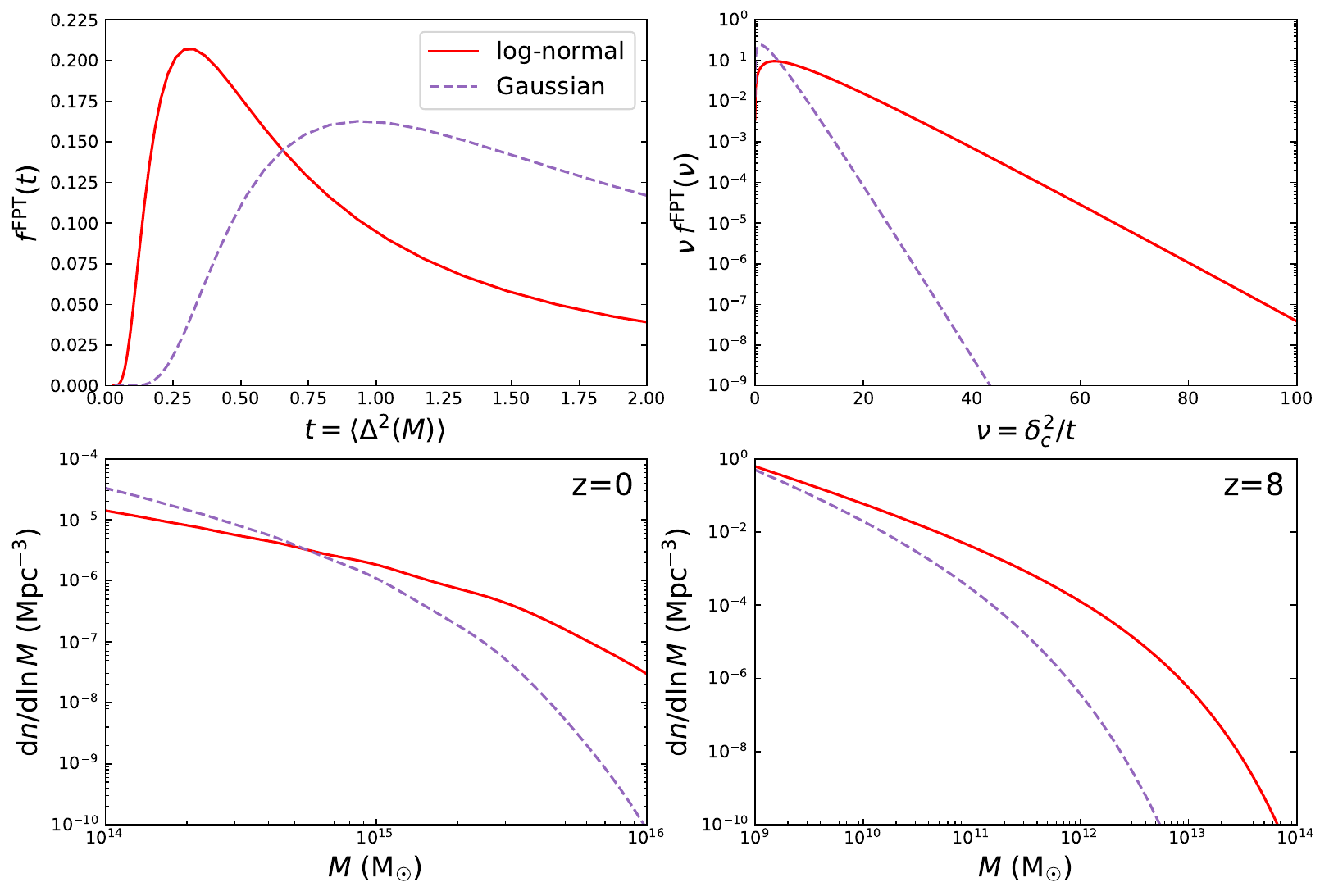}
    \caption{{\it Top left:} First-passage time distribution \eqref{eq:pdf-moving-barrier} for the moving barrier $c(t_y)=\ln(1+\delta_c)+t_y/2$. {\it Top right:} Same distribution using the customary variable $\nu \equiv \delta_c^2 /t$, i.e. $\nu f\fpt(\nu) = \nu (\dd t/\dd\nu)f\fpt(t)$. {\it Bottom:} Corresponding halo mass functions \eqref{eq:HMF} at redshifts $z=0$ ({\it left}) and $z=8$ ({\it right}). We have kept the same axis bounds as \fig\ref{fig:1barrier} for easier comparison, even if this result does not rely on the assumptions of the LDP.}
    \label{fig:lognormal}
\end{figure}

For linearly moving barriers of Gaussian random walks, exact expressions for the first crossing time distributions exist~\cite{Sheth:2001dp_tormen_ellips,DeSimone:2010mu_moving_bar};
\begin{equation}
    \label{eq:pdf-moving-barrier}
    p_{T_{c(t_y)}}(t_y) = f\fpt_y(t_y) = \frac{c(\delta_c, 0)}{\sqrt{2\pi} t_y^{3/2}} \exp\left(- \frac{c(\delta_c,t_y)^2}{2t_{y}}\right)
\end{equation}
We want to connect the crossing time for $Y$ to the halo mass function that, by contrast, depends on the crossing time for $\De$. As mentioned in Section \ref{sub:excursionset}, the correspondence between the coarse-graining scale $R$ and the random walk time variable $t$ is one-to-one. By writing $t_y = \psi_1(R)$  and $t_\delta = \psi_2(R)$, we have $t_\delta = \psi_2\circ \psi_1^{-1} (t_y)$. The first-passage time $T^\delta_b$ for $\De$ at barrier $\delta_c =b=1,686$ is then given by
\begin{equation}
    T_b^\delta = \psi_2\circ \psi_1^{-1}(T_{c(b, t_y)}^y) \mperiod
\end{equation}
From this it follows $f\fpt_\delta(t_\delta) = (\dd t_y/\dd t_\delta) f\fpt_y(t_y)$. We can plug this into \Eq\eqref{eq:HMF}. Using \Eq\eqref{eq:ty-vs-tdelta} we obtain a corrected formula for the halo mass function,
\begin{equation}
    \frac{\dd n}{\dd\ln M} = \frac{\bar{\rho}_{0,m}}{M} \frac{\dd t_y}{\dd t_\delta} f\fpt_y(\psi_1(R))\times \frac{1}{3}\left|\frac{\dd t_\de}{\dd \ln R}\right| \mperiod
\end{equation}
This new halo mass function is shown in \fig\ref{fig:lognormal}.

Here we have considered the log-normal case. We could also consider the possibility that $Y$ would not be Gaussian but, e.g., has arbitrary exponential tails in the form of \Eq\eqref{eq:pdt-with-polynom} However, in this case the motion 
of the barrier $c(t_y)$ would no longer be simply linear in $t_y$ and we would have to make approximations to describe it, see~\cite{Sheth:2001dp_tormen_ellips,DeSimone:2010mu_moving_bar}. On the other hand, we could consider more general functions $Y=F(1+\De)$ that are symmetric for arguments $1+\de\gs 0$. We leave these extensions for future work.

\section{Abundance of cosmic voids}
\label{sec:voids}

As a second application of the large deviation principle, we turn our attention to underdensities, i.e. the seeds of cosmic voids. This was our original motivation to bring new theoretical predictions, as void theory still needs to be consolidated \cite{Pisani:2019cvo_whitepaper}. We present here an updated, non-perturbative  prediction of the void size function (VSF) within the excursion set formalism, for  density contrast distributed with the exponential tails \eqref{eq:pdt-with-polynom}. Although publications discussing the effect of perturbative non-Gaussianities on the VSF can be found~\cite{DAmico:2010dwy_void_png}, to our knowledge an explicit computation using strongly non-Gaussian tails has never been performed.

\begin{figure}
    \centering
    \includegraphics[width=\linewidth]{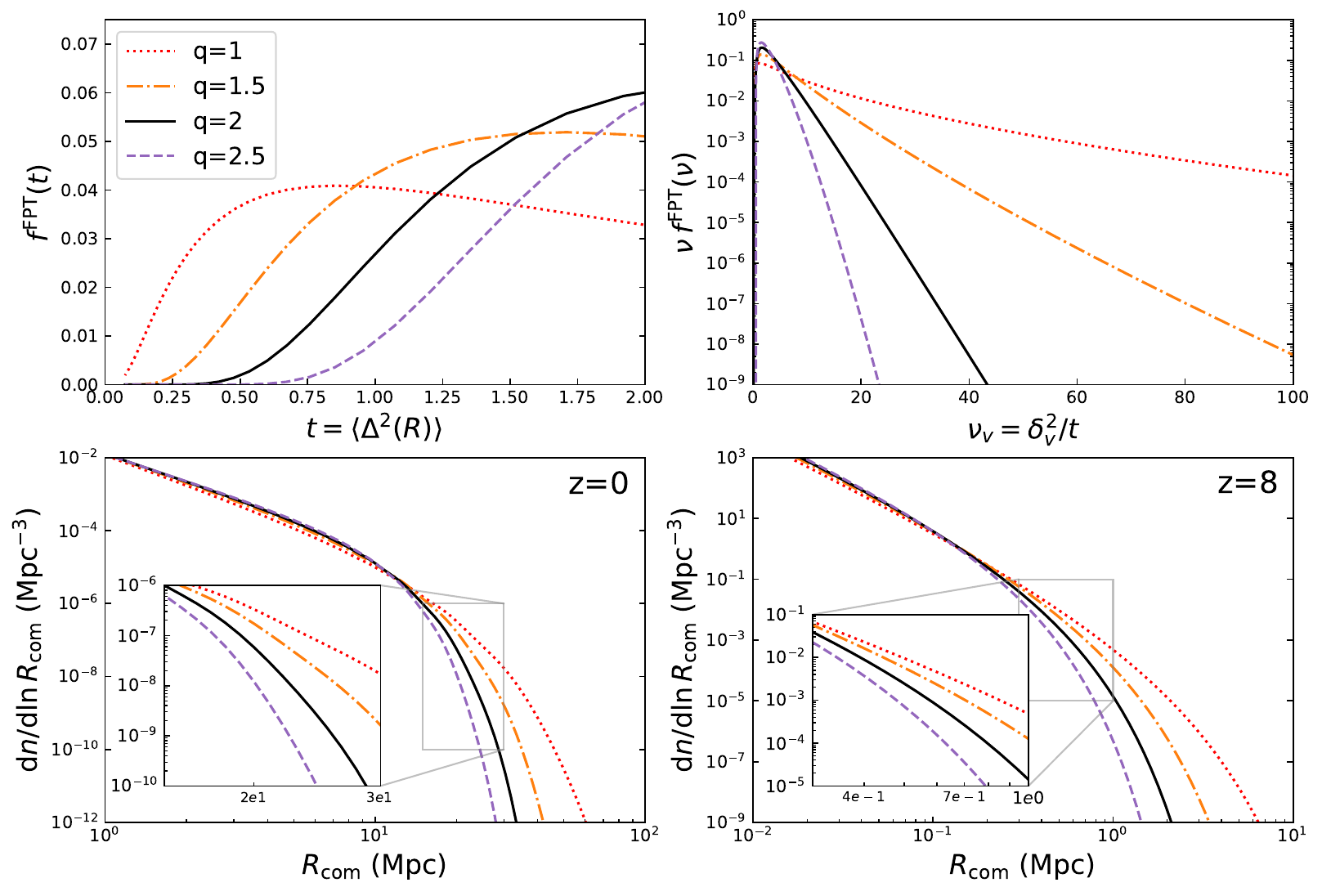}
    \caption{{\it Top left:} Non-gaussian first-passage time distribution \eqref{eq:fv-main-text} at $a=\delta_v = -2.72$ before $b=\delta_c = 1.686$ (see the text), for various values of $q$, and $\alpha = (q-2)/2$. {\it Top right:} Same distribution using the customary variable $\nu_v \equiv \delta_v^2 /t$, i.e. $\nu_v f\fpt(\nu_v) = \nu_v (\dd t/\dd\nu_v)f\fpt(t)$. {\it Bottom:} Corresponding void size function \eqref{eq:vsf} at redshifts $z=0$ ({\it left}) and $z=8$ ({\it right}). The black line $q=2$ corresponds to the standard Gaussian distribution. Horizontal axis bounds have been chosen to match the regime where the LDP is applicable, $t < \delta_c^2$, see the caption of \fig\ref{fig:variance-to-R}.
    \label{fig:2barriers}}
\end{figure}

In the standard picture, void formation is modeled as the passage of the random walk under a negative barrier, $a=\delta_v = -2.72<0$ for the spherical void model in linear perturbation theory \cite{Sheth:2003py_seminal}. Obviously, the true non-linear density contrast associated to this linear theory threshold always remains larger than $-1$. The VSF is defined in a similar fashion as the halo mass function \eqref{eq:HMF} and provides the comoving number density of voids of comoving size $R\subsc{com}$ \cite{DAmico:2010dwy_void_png},
\begin{equation}
\label{eq:vsf}
    \frac{\dd n(z)}{\dd\ln R\subsc{com}} = \frac{V(R)}{V(R\subsc{com})}\frac{3}{4\pi (cR)^3}f\fpt_v(t)\abs{\frac{\dd t(z)}{\dd \ln R}}_{R=R\subsc{com}/1.7}\mperiod
\end{equation}
Here, $V$ is defined by \Eq\eqref{eq:R-vs-M}. Let us provide a few explanations. First, the link between the comoving radius $R\subsc{com}$ and the linear theory radius $R$ is found to be, for the spherical void model, $R\subsc{com} \simeq 1.7 R$ \cite{1992_blumenthal_voids_seminal,DAmico:2010dwy_void_png}. Next, we include a volume correction factor $V(R)/V(R\subsc{com})$ that is based on the so-called Vdn model \cite{Jennings:2013nsa_Vdn_model,Verza:2019tvg_pisani_for_Vdn}. This factor was not present in seminal papers on void theory \cite{Sheth:2003py_seminal} and has been proposed in more recent developments \cite{Jennings:2013nsa_Vdn_model}. Because cosmic voids can merge as they grow during the expansion of the Universe, their number density is not conserved. Such an effect was not accounted for in early works. The Vdn model alleviates this issue by identifying a better conserved quantity (namely, $V\dd n$ rather than $n$), and has been shown to improve accordance between theory and simulations. In the spherical collapse model case that we consider, this introduces a correction $1/1.7^3$ to the VSF. Last but not least, the definition of $f\fpt_v$ is theoretically more involved than $f\fpt$. As was originally pointed out in Ref. \cite{Sheth:2003py_seminal}, accounting for the ``void-in-cloud" effect imposes to exclude random walks passing above $b=\delta_c = 1.686$ before $a=\delta_v=-2.72$, as these end up in collapsed objects rather than voids. The definition of $f\fpt_v$ is thus
\begin{equation}
    f\fpt_v(t) \dd t = \Pro(T_a \in \interval{[}{t}{t+\dd t}{[}\ \text{and} \ T_a < T_b) \mperiod
\end{equation}
In their seminal paper~\cite{Sheth:2003py_seminal}, the authors compute $f\fpt_v$ in the Gaussian case, $q=2$ and $\al=0$, by using a connection between the Laplace transforms of $f\fpt$ and $f\fpt_v$. We now extend this methodology to all values of $q\gs 1$, within the assumptions of the LDP. The derivation is somewhat involved, so we only sketch the main steps here and refer the reader to Appendix \ref{appx:sub:2-barriers} for the complete solution. We denote $\ell_1(s,\delta_c)$ and $\ell_2(s,\delta_v,\delta_c)$ the respective Laplace transforms of $f\fpt(t,\delta_c)$ and $f\fpt_v(t,\delta_v,\delta_c)$. They follow the relationship \cite{Sheth:2003py_seminal}
\begin{equation}
\label{eq:1-to-2-barriers-maintxt}
    \ell_2(s,\delta_v,\delta_c) = \frac{\ell_1(s,\delta_v)-\ell_1(s,\de_c)\ell_1(s,\de_c -\de_v)}{1-\ell_1(s,\de_c-\de_v)^2}\mcomma
\end{equation}
a property relying on Markovianity but, crucially, not on Gaussianity. Considering the single-barrier FPT distribution $f\fpt(t,\delta_c)$ to be given by \Eq\eqref{eq:pdt-with-polynom}, we find a simple expression for its Laplace transform at large arguments,
\begin{equation}
\label{eq:1bar-laplace-maintxt}
     \ell_1(s,\delta_c) \underset{s\to +\infty}{\simeq} \sqrt{Q}\exp(- \frac{q+2}{2} (2s \delta_c^2 \gamma^2 /q)^{Q/2}) \mcomma
\end{equation}
where we have defined 
\begin{equation}
    Q\equiv 2q/(q+2) \mperiod
\end{equation}
In the process of obtaining \Eq\eqref{eq:1bar-laplace-maintxt} we have exploited the freedom of the exponent $\al$ in \Eq\eqref{eq:pdt-with-polynom}, unconstrained by the LDP, to set $\al=(q-2)/2>-1$ (note that for $q=2$ this gives $\alpha=0$, as in the original Ref. \cite{Sheth:2003py_seminal}; also, $\gamma = (\Gamma(1/2 + 2/q)/\sqrt{\pi})^{1/2}$ now). This allows us to simplify Laplace transform expressions without affecting the exponential behavior of the distribution. We next insert this expression into \Eq\eqref{eq:1-to-2-barriers-maintxt}, expand the denominator into series and proceed to reconstruct the double barrier probability $f\fpt_v$ from its Laplace transform (c.f. Appendix \ref{appx:sub:2-barriers} for details). We eventually find 
\begin{align}
     f\fpt_v(t)  = \frac{q}{2\sqrt{\pi} t} &\sum_{n\gs 0}\left[Q^{n} \left(\frac{c_n \gamma}{\sqrt{t}}\right)^{q/2}\exp\left(-\left(\frac{c_n \gamma}{\sqrt{t}}\right)^{q}\right)\right. \nonumber \\
    &\qquad\left.-\ Q^{n+1/2}\left(\frac{d_n \gamma}{\sqrt{t}}\right)^{q/2}\exp\left(-\left(\frac{d_n \gamma}{\sqrt{t}}\right)^{q}\right)\right]\mperiod
    \label{eq:fv-main-text}
\end{align}
where we have introduced new barrier heights,
\begin{equation}
   \forall n\gs0,\ c_n \equiv (\abs{a}^{Q}+2n(b-a)^{Q})^{1/Q}\quad,\quad d_n \equiv (b^{Q}+(2n+1)(b-a)^{Q})^{1/Q}\mperiod
\end{equation}
This constitutes our main new result. \Eq\eqref{eq:fv-main-text} also admits the following expression,
\begin{equation} \label{e:fFPTvoid}
    f\fpt_v(t) = \sum_{n\gs 0} \left[Q^n f\fpt(t\mid b=c_n, \al=\frac{q}{2}-1) - Q^{n+1/2} f\fpt(t\mid b=d_n, \al=\frac{q}{2}-1)\mcomma\right]
\end{equation}
where $f\fpt(t)$ is the halo barrier function given in \Eq\eqref{eq:1st-passage-expo-tail} with corresponding parameters. The sum in \Eq\eqref{e:fFPTvoid} converges very rapidly as the barriers $c_n$ and $d_n$ rapidly grow very large. This expansion reduces to the expression found in  Ref. \cite{DAmico:2010dwy_void_png} in the case $q=2$ and $\al=0$, itself consistent with other works \cite{Sheth:2003py_seminal,Lam:2009nd_before_DAmico}. We present it in \fig\ref{fig:2barriers} together with the corresponding VSF \eqref{eq:vsf}. 

By the nature of the LDP, \Eq\eqref{eq:fv-main-text} is valid for large fluctuations $\left|\de\right| \gg \sqrt{\avg{\De^2}}=\sqrt{t}$ or,  equivalently, for  small $t$. Let us remark that by Gambler's ruin property, $f\fpt_v$ should in principle be normalized to $\int_0^\infty f\fpt_v(t)\dd t = b/(b-a)$ \cite{Sheth:2003py_seminal}. However, the expression we have found cannot be used for arbitrary large values of $t$ that are outside the LDP regime. Only in the exceptional case $q=2$, where the expression obtained with the LDP matches the exact result at all $t$'s, can this normalization be readily satisfied.

As in the case of over densities, the difference between this new prediction and the Gaussian case is more pronounced in the small variance limit $t\to 0$. This is reassuring: the validity domain of the LDP is also where the effect that it predicts is the strongest. However, as pointed out in Ref. \cite{DAmico:2010dwy_void_png}, this is also the limit in which solutions to the two-barrier and one-barrier problems become almost indistinguishable, as it becomes increasingly unlikely that the walk reaches $\delta_c>0$ {\it and} comes back to $\de_v<0$ in a very small ``time" interval. This is reflected in the convergence speed of the series \eqref{eq:fv-main-text}, which becomes faster for $t\to 0$. Also in this limit the smallest barrier, $c_0$, dominates the series, and the associated term is exactly the one barrier result \eqref{eq:1st-passage-expo-tail}.  

\section{Discussion and conclusion}
\label{sec:opening}

In this work we have applied the theory of large deviations to compute the theoretical abundance of large halos and voids from non-Gaussian initial conditions, motivated by predictions coming from inflationary models. We have seen that the large deviation principle is well adapted to the quantitative description of large non-Gaussianities that are still escaping constraints in the tails of the distribution, despite the stringent limits e.g. on $f\subsc{NL}$ from the analysis on CMB scales \cite{Planck:2019kim_PNG_constr}. 
We have shown that the LDP can provide analytical solutions to problems where traditional probability theory either fails or is hardly tractable. This large deviation approach is accurate when the following (intertwined) conditions are met: {\it (i)} large, rare fluctuations (compared to the typical variance); {\it (ii)} large scales (where the variance is small). At redshift zero, this corresponds to sizes larger than $R\sim\, \text{few Mpc}$ or masses larger than $M\sim (10^{13}-10^{14})\solarmass$. At higher redshifts the validity domain extends to smaller scales because the variance decreases, $\avg{\De^2} \propto D^2(z)$ where $D(z)$ is the linear growth factor \cite{Eisenstein:1997jh_power_spectrum}. For instance, at $z=8$ the LDP is applicable for $R> 10^{-2} \mathrm{Mpc}$ and $M>10^7 \solarmass$. 

Reassuringly, the deviations from Gaussianity predicted using the LDP are the strongest precisely in the regime where its validity is most robust. A contrario, our method is not directly applicable to small fluctuations $\abs{\de} \ll \sqrt{\avg{\de^2}}$. But we expect their distribution to be close to Gaussianity, where exact results are known. In a similar fashion, the LDP is also insensitive to perturbative non-Gaussianities of the type $\delta=\delta_G +f\subsc{NL}(\delta_G^2-\avg{\delta_G^2})$ with $\delta_G$ a Gaussian field. Such an expansion intrinsically assumes the field $\delta_G(x)$ to take small enough values -- else, more terms in this $\delta_G$-expansion are needed. These small values do not probe the tail of the distribution.\footnote{In a more realistic setup, one should consider the distribution to be (almost) Gaussian for small fluctuations, and to present exponential tails in the large fluctuation limit, schematically $p(\delta) \propto \exp(-A(\delta/\sig)^2) + \exp(-B \abs{\delta}/\sig)$. Since we are only analyzing the behavior of large fluctuations, this correction is not very significant to our purposes.} The LDP is also insensitive to subexponential corrections to the distributions, like polynomial prefactors. In that sense, the LDP explored in this work, is complementary to the perturbative developments found in a vast body of literature.

We have then combined the LDP with the excursion set formalism to study the effect of exponential tails, as a typical example of strongly non-Gaussian distributions. We stress that the excursion set formalism only requires the knowledge of the linear theory. That is, the non-Gaussianities that are accounted for in this work come from early Universe physics and are not seeded by late non-linear effects like non-linear clustering.

For (symmetric) exponential tails \eqref{eq:pdt-with-polynom} ($1\ls q<2$), both large halos and voids are more probable than in the Gaussian case. The difference is most pronounced at high redshifts. For $z=8$, haloes of 
$10^{12}\solarmass$ are about 100 times more probable than for a Gaussian distribution. For  
$10^{13}\solarmass$ this factor raises even to $10^5$. Therefore, exponential tails (or a log-normal distribution) may in principle explain the large galaxies observed at high redshift by the James Webb Space Telescope~\cite{JWST:2023_little_red_dots}. The analysis~\cite{Boylan-Kolchin:2022kae_little_red_dots} that found these high redshift galaxies to be incompatible with $\La$CDM fully relied on the Press-Schechter abundance formula that of course only holds for Gaussian distributions. For these distributions the abundance of large voids is also significantly enhanced. At $z=8$ the abundance of voids of $2 \mathrm{Mpc}$ is enhanced by $10^4$ and the density of larger voids is enhanced by even more. At $z=0$, this enhancement factor even reaches $10^5$ for voids of size $40 \mathrm{Mpc}$, and increases for even larger voids. Another consequence of these non-Gaussian tails is that the effective threshold barrier for halo formation increases as $q$ is lowered. We also observe that  on large scales decreasing the value of $q$ has a similar effect on the mass function as increasing the value of $\sigma_8$, as is commonly done in halo counting fits \cite{Mo_van_den_Bosch_White_2010_book}. Small $q$'s however lead to a depletion of the mass function at small masses, unlike the increase $\sigma_8$ that boosts the mass function on all scales \cite{moyeranin:tel-05496942_alice_mass_function}. It would then be interesting to compare these effects and see what degeneracies may be alleviated by introducing $q$ as a free parameter in the initial conditions. 

Finally, we have explored the case of a log-normal distribution where over- and under-densities are no longer symmetric. This is more realistic as $\de$ cannot be smaller than $-1$ while it can, in principle have arbitrary positive values. With a log-normal distribution large halos are much more probable than in the Gaussian case, to the detriment of large under-densities. At $z=0$ halos of mass $10^{16}\solarmass$ are about $10^3$ more probable for a log-normal distribution than in the Gaussian case. At $z=8$ the increase  for $10^{13}\solarmass$ halos even becomes a factor of $10^5$.

This work is intended as a proof of concept, exposing the power of large deviation theory to establish a connection between inflation models and late Universe observables. We identify two clear directions which  can be pursued within this program. First, our results need to be applied to exponential-like tails commonly obtained in the inflationary $\delta N$-formalism \cite{Pi:2022ysn_deltaN_pdf,Cruces:2025typ_deltaN_pdf,Coulton:2024vot_png_vs_data,Gow:2022jfb_nonpert_PBH_1pt_func}. These are usually derived in real space, for which the correlations are non-trivial. Their connection to the Fourier space distribution where these correlations, in a first approximation, can be neglected, needs to be studied. On the other side, we want to compare our predictions to simulations -- and eventually observations, especially in the case of voids. Large voids are notoriously difficult to identify within datasets \cite{Contarini:2026yfv_pisani_recent_review}, hence a better connection between the theoretical prediction and measurements would certainly be mutually beneficial. The realm of applications of the LDP in cosmology is still vastly uncharted, and we shall continue to explore its rich phenomenology in future works.

\section*{Acknowledgments}

M.T. is profoundly indebted to Dylan Thévenet, without whom this work would never have come to fruition. M.T. thanks Julien Lavalle and Lucas Pinol for insightful discussions in the great venue of the Palais des Papes d'Avignon. The authors thank Céline Combet, Rosa Malandrino, Azadeh Moradinezhad,  Alice Pisani and Vincent Vennin for helping strengthen the scope of this article.

\appendix

\section{Consequences of statistical homogeneity}
\label{appx:homo}

The statements presented throughout this article rely on several key properties concerning the independence of random variables. We split here properties that are only based on statistical homogeneity\footnote{Interestingly, as one may convince themselves by reading the proofs, they do not require isotropy but solely homogeneity.} of the Universe, from those that rely on fluctuations following joint Gaussian distributions. See also Ref. \cite{Fan:1995aq_bardeen_CLT_cosmo} on the matter.

Let $\De:\R^3 \to \R$ a real random field  and $\hade:\R^3 \to \R$ its Fourier transform. It is assumed that $\Delta$ is statistically homogeneous, that is, $\forall a\in\R^3$, $\De(\cdot)$ and $\De(\cdot+a)$ have the same probability distribution. Then the following are true:
\begin{enumerate}[label={\it (\roman*)}]
    \item $\forall k \in\R^3$, $\arg(\hade_k)$ is uniformly distributed on $\interval{[}{0}{2\pi}{[}$;
    \item $\forall k \in\R^3$, $\arg(\hade_k)$ and $|\hade_k|$ are independent;\footnote{\label{foot8}Note however that ${\rm Re}(\hade_k)$ and ${\rm Im}(\hade_k)$ are generally not independent. The Herschel-Maxwell theorem states that independent components of a   random vector field with rotational symmetry are necessarily Gaussian~\cite{mukherjee2017proofherschelmaxwelltheoremusing}. Therefore the independence of ${\rm Re}(\hade_k)$ and ${\rm Im}(\hade_k)$ together with point {\it (i)} above would imply that $\hade_k$ is necessarily Gaussian. One can only show in general that these are uncorrelated rather than independent, $\avg{{\rm Re}(\hade_k){\rm \Im}(\hade_k)}=0$.}
    \item if the set of random variables $\{\hade_k\}_k$ is jointly Gaussian distributed, then the random variables $\{\hade_k, k\in \Ralft\}$ are mutually independent.
\end{enumerate}

\begin{figure}
    \centering
\includegraphics[width=0.4\linewidth]{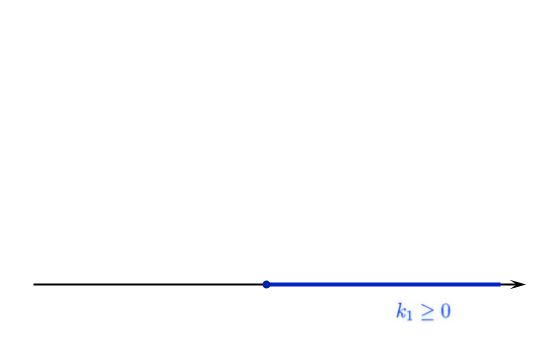}
    \includegraphics[width=0.4\linewidth]{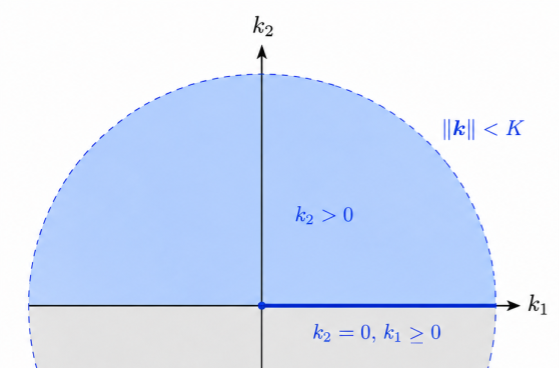}
    \includegraphics[width=0.55\linewidth]{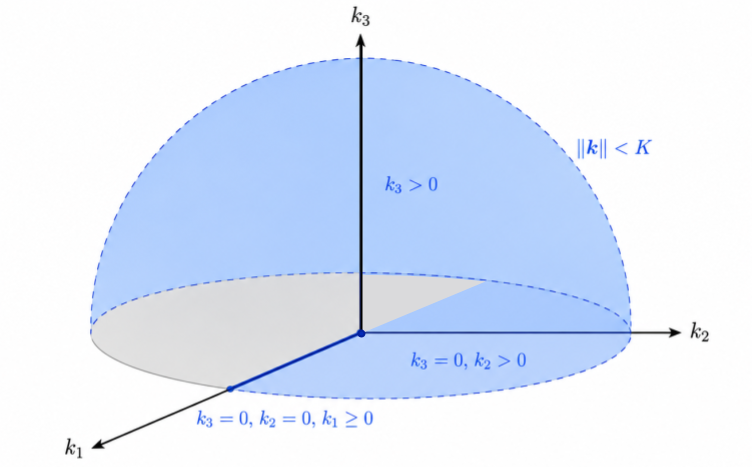}
    \caption{$\Ralf^d$ for $d=1,2,3$, limited to $\norm{k} < K = R^{-1}$. Any closed polygon of vectors in $\R^d$ must have at least one (oriented) edge that does not belong to $\Ralf^d$.}
    \label{fig:halfspace}
\end{figure}

Let us recall the definition of $\Ralft$:
\begin{align}
    \Ralf^0 &= \{0\}  \nonumber \intertext{and}
\label{eq-appx:def-halfspace}
    \forall d\in\N_0,\  \Ralf^{d+1} &= \left\{\begin{pmatrix}
        k_1 \\ \vdots \\ k_{d+1} 
    \end{pmatrix} \Big|\; k_{d+1} > 0 \ \text{or}\ \left(k_{d+1}=0\ \text{and}\ \begin{pmatrix}
        k_1 \\ \vdots \\ k_{d} 
    \end{pmatrix}\in \Ralf^d \right)\right\} \mperiod \qquad
\end{align}
By construction, $\Ralf^d$ is such that 
\begin{itemize}
    \item no two vectors $k$ and $-k$ simultaneously belong to $\Ralf^d$ unless $k=0$ (and in fact $\hade_0 =0$ from $\avg{\De(x)} = 0)$;
    \item no closed polygon can be formed with vectors of $\Ralf^d$ (see \fig\ref{fig:halfspace}). 
\end{itemize}

Let us now prove the aforementioned statements.

\begin{enumerate}[label={\it (\roman*)}]
    \item Let $\phi\in\R$. We write
    \begin{equation}
        e^{-i\phi} \hade_k = \int \dd^d x e^{-i(k\cdot x + \phi)} \De(x) = \int \dd^d y e^{-i k \cdot y} \Delta'(y) \mcomma
    \end{equation}
where $y=x+\phi k/\Vert k \Vert^2$ and $\Delta'(y) = \Delta(y-\phi k/\Vert k \Vert^2)$. By statistical homogeneity, $\Delta'$ and $\Delta$ are identically distributed, hence also $e^{-i\phi} \hade_k$ and $\hade_k$ have the same distribution (rotational symmetry).

Let us now write $Z=\hade_k = U e^{i\Th}$ with probability density $p_Z(u,\th)$. From rotational symmetry it follows that $\forall\phi\in\R$, $p_Z(u,\th)=p_Z(u,\th+\phi)$, hence $p_Z(u,\th) = f(u)$ is independent of $\th$. Using the normalization of $p_Z$, one obtains $\int_0^\infty f(u)u \dd u = 1/(2\pi)$. The marginalized density for $\Th$ hence reads $p_\Th(\th) = \int_0^\infty p_Z(u,\th) u \dd u = 1/(2\pi)$, so $\Th$ is uniformly distributed.

\item Using the previous proof, the marginalized density for $U$ reads $p_U(u) = 2\pi u f(u)$. Thus, $p_{(U,\Th)}(u,\th) = u p_Z(u,\th) = u f(u) = \frac{1}{2\pi} p_U(u) = p_U(u) \times p_\Th(\th)$, showing that $U$ and $\Th$ are independent.

\item By definition, $\{\hade_k\}$ are jointly (complex) Gaussian if for any $r\in\N$, and $k_1,\dots,k_r\in \R^3$ two-by-two distinct, the joint distribution of $\bm{\De} =(\hade_{k_1},\dots,\hade_{k_r})$ takes the form
\begin{equation}
\label{eq-appx:def-joint-gauss}
    p\supsc{joint}_{\bm{\De}}(\bm{\de)} = p_{\bm{\De}}(\de_{k_1},\dots,\delta_{k_r}) = \frac{1}{\pi^r (\det \Sigma\det C)^{1/2}}\exp[-\frac{1}{2} \begin{pmatrix}
        \bm{\delta} \\ \bm{\delta}^*
    \end{pmatrix}^\dagger \begin{pmatrix}
        \Sigma & C \\ 
        C^* & \Sigma^*
\end{pmatrix}^{-1}\begin{pmatrix}
        \bm{\delta} \\ \bm{\delta}^*
    \end{pmatrix}] \mcomma
\end{equation}
where $\Sigma = (\Sigma_{kl})_{kl} = \avg{\bm{\Delta} \bm{\Delta}^\dagger}$  and $C=(C_{kl})_{kl} = \avg{\bm{\Delta} \bm{\Delta}\!\supsc{T}}$ are the covariance and pseudo-covariance matrices. From statistical homogeneity, it is well-known that $\avg{\hade_k \hade_l}\propto \delta_D^{(3)}(k+l)$. Therefore, if we now restrict the modes $k_1,\dots,k_r\in \Ralft$ to be in the half-space \eqref{eq-appx:def-halfspace}, we find $C= 0$ (observing that $\hade_0 = \int \dd^3 x \De(x) =0
$) and $\Sigma$ is diagonal. The joint probability factorizes,
\begin{align}
     p\supsc{joint}_{\bm{\De}}(\bm{\de)} &= \prod_{i=1}^r  \frac{1}{\pi \avg{|\hade_{k_i}|^2}}\exp(-\frac{\delta_{k_i}^*\delta_{k_i}}{\avg{|\hade_{k_i}|^2}}) \\
     &= \prod_{i=1}^r p_{\De_{k_i}}(\de_{k_i})\mcomma
\end{align}
so modes $\hade_k, k\in\Ralft$ belonging to the half-space form a set of mutually independent variables. Note also that, for these same modes, from circular invariance correlators at all orders vanish,
\begin{equation}
\avg{\hade_{k_1}\cdots \hade_{k_n}} =0    \mcomma
\end{equation}
but because the variables are complex, without the assumption of Gaussianity this enough is not enough to  show mutual independence.

We also emphasize that neither independence, nor the fact that the distribution is entirely determined by the two-point correlation, nor Wick theorem apply if the $\De_k$'s are each individually sampled from a Gaussian distribution, but not jointly Gaussian (in the meaning of \Eq\eqref{eq-appx:def-joint-gauss}. A typical counter-example is to consider a Gaussian variable $X \hookrightarrow \mathcal{N}(0,\sigma^2)$, another random variable $S=\pm 1$ with equal probability $1/2$, independent from $X$, and $Y=SX$. Like $X$, $Y$ is distributed according to the distribution $\mathcal{N}(0,\sigma^2)$, and $\avg{X^2 Y^2}=\avg{X^4 S^2} =3 \avg{X^2}^2$. However, if Wick's theorem would apply we would find $\avg{X^2 Y^2} = \avg{X^2}\avg{Y^2} + 2 \avg{XY}^2 = \avg{X^2}^2$, which is not correct here. Moreover, despite $X$ and $Y$ being Gaussian the sum $X+Y$ is not a Gaussian, since it takes values in the zero-measure set $\{0\}$ with probability $1/2$.
\end{enumerate}

\section{The Large Deviation Principle}
\label{appx:LDP}

\subsection{Basics}
\label{appx:sub:basics}

The notion of large deviations considered in this article is a  simplified version of the general large deviation principle (LDP) that is sufficient for our purpose. We point the interested reader to Ref.~ \cite{dembo2009_zeitouni_book_LDP}, one of the most recognized and exhaustive references on the subject, while Refs.~\cite{Touchette_2009_review,Burenev_2025_touchette_recent} review the LDP within physical contexts in a  language more accessible to physicists.

\begin{itemize}
    \item {\it Large deviation principle.} 
    Consider a set of complex random variables $(X_\eps)_\eps$, indexed by $\eps >0$. The family $(X_\eps)_\eps$ is said to satisfy a large deviation principle (LDP) for the {\it rate function} $I:\C\to \R$ if for any measurable set $B\subset \C$ (or $B \subset \R$ if the $X_\eps$'s are real),
\begin{equation}
\label{eq:def-LDP-appx}
    \lim_{\eps\to 0} \eps \ln\Pro(X_\eps\in B) = - \inf_{z\in B}  I(z)
    \mperiod
\end{equation}
In our case the random variables are always circularly symmetric, so\\ $\inf_{z\in B}I(z) = \inf_{r\in B \cap \R_+} I(r) $ and the LDP can be intuitevely interpreted as saying 
\begin{equation}
    \Pro(X_\eps\in B) \underset{\eps\to 0}{\asymp} \exp(-\inf _{r\in B\cap \R_+} I(r) / \eps) \mcomma
\end{equation}
where ``$\asymp$" is an equivalence up to subexponential factors, in the meaning of \Eq\eqref{eq:def-LDP-appx}. An equivalent definition can be made at the level of the probability density. Assuming again circular symmetry and setting $B=C(r) = \{z\in\C\ | \ \abs{z}\in\interval{[}{r}{r+\dd r}{[}\}$, $\Pro(X_\eps \in C(r)) = p_{X_\eps}(r) \dd r$ and 
\begin{equation}
    p_{X_\eps}(r) \underset{\eps\to 0}{\asymp} \exp(-I(r)/\eps) \mperiod
\end{equation}
In other words, when $(X_\eps)_\eps$ satisfies the LDP, the probability density of $X_\eps$ decays exponentially in $1/\eps$. When $\eps$ corresponds to the typical variance of $X_\eps$, this can be used to study the tail of the distribution, $X_\eps \gg \avg{\abs{X_\eps}}$.

\item {\it Contraction principle.} Let $F:\C^d\to\C^d$ be a continuous map and $(X_\eps)_\eps,(Y_\eps)_\eps \in \R^d$ be two families of random variables such that $Y_\eps=F(X_\eps)$. We would like to know the distribution of $Y_\eps$ depending on the of $X_\eps$. Under some technical assumptions (see §4.2.1 of Ref. \cite{dembo2009_zeitouni_book_LDP}), if $(X_\eps)_\eps$ satisfies a LDP with rate function $I_X(x)$, then $(Y_\eps)_\eps$ also satisfies a LDP with rate  function provided by
\begin{equation}
\label{eq-appx:contraction-prin}
    I_Y(y) = \inf\{I_X(x)\ ,\ x\in \C^d \ \text{such that}\ F(x) = y\}\mperiod
\end{equation}

\item {\it Contraction principle (bis).} It is worth stating a version of theorem \eqref{eq-appx:contraction-prin} whenever the components of $X_\eps$ are mutually independent. Let $(X^{(1)}_\eps)_\eps,(X^{(2)}_\eps)_\eps \in \R$ be two independent sets of random variables. Assume that they both satisfy a LDP with rate functions $I_1(x_1)$ and $I_2(x_2)$. Under additional technical constraints (see §4.2.7 of Ref. \cite{dembo2009_zeitouni_book_LDP}), we can show that for any continuous function $F:\C^2\to \C$ the random variable $Y_\eps = F(X^{(1)}_\eps,X^{(2)}_\eps)$, satisfies a LDP with rate function 
\begin{equation}
\label{eq:optimization}
    I_Y(y) = \inf_{\{(x_1,x_2)|F(x_1,x_2)=y\}}I_1(x_1)+I_2(x_2) \mperiod
\end{equation}
This generalizes easily to more than two sets of independent random variables. This powerful theorem allows a reformulation of the original problem: instead of searching the probability density of $F(X^{(1)},X^{(2)})$, which for a general $F$ is a very complicated problem, one can rather solve an optimization problem to get information on its tail -- a task that tends to be much simpler. Another strength of this approach is that $F$ needs not to be a bijection: one may obtain the rate function of a single random variable that is built from arbitrary many others.

\item {\it Application of the Gärtner-Ellis lemma.} In Section \ref{sub:expo-tails} we  consider probability densities of the form
\begin{equation}
    p(\delta) = Q(\delta) \exp(-\abs{\de}^q /\eps)\mcomma
\end{equation}
where $Q(\delta)$ is any function with a typical power-law behavior (see the text). Their rate functions can readily be obtained from the definition \eqref{eq:def-LDP-appx}, 
\begin{equation}
    I(\delta) = \abs{\delta}^q \mperiod
\end{equation}
Here we rederive this result from the Gärtner-Ellis lemma, as the latter may be more easily generalizable. The lemma reads \cite{Touchette_2009_review}
\begin{equation}
    I(\delta) = \sup_{y}\{y\cdot\delta -\Lambda(y)\}\quad,\quad \Lambda(y)\equiv \lim_{\eps\to 0} \eps\ln\avg{e^{(\Delta\cdot y) /\eps}} \mperiod
\end{equation}
Be $y_*$ the solution of $\nabla_y(y\cdot \delta - \Lambda(y))=0$, which leads to $\delta = \Lambda'(y_*)$. Using Laplace's method we then evaluate
\begin{align}
   \delta =  \Lambda'(y_*) &=\lim_{\eps\to 0}\frac{\avg{\Delta e^{(\Delta\cdot y_*) /\eps}}}{\avg{e^{(\Delta\cdot y_*) /\eps}}} \\
    &= \lim_{\eps\to 0}\frac{\int u Q(u) e^{f(u)/\eps}\dd u}{\int Q(u) e^{f(u)/\eps}\dd u}\quad, \quad f(u) \equiv y_*\cdot u - \abs{u}^q \\
    &\underset{\eps\to0}{\simeq} u_* \mcomma
\end{align}
where $u_*$ solves $f'(u_*)=0$ i.e. $q\abs{u_*}^{q-1} = y_*$. With a second Laplace method one finds $\avg{e^{(\De\cdot y_*)/\eps} }\simeq \exp( f(u_*)/\eps)$, hence $\Lambda(y_*) \simeq f(u_*) = q \abs{u_*}^q - \abs{u_*}^q = (q-1) \abs{\delta}^q$. Finally, $I(\delta) = q\abs{\delta}^{q} - (q-1)\abs{\delta}^q = \abs{\delta}^q$. \hfill $\square$ 

This derivation holds whether $\De$ is a scalar or a multi-component vector.

\end{itemize}

\subsection{A lemma for optimization problems}
\label{appx:sub:lemma}

We prove here our result \eqref{eq:result-of-LDP} that allows to apply the LDP in the context of random Fourier transform. Fix $q>0$ and $K\subset \R^3$ a measurable set. We denote for complex-valued functions $f,g:K\to \C$
\begin{align}
\label{eq:def-qnorm}
   \forall f\in L^q(K),\ \Vert f\Vert_{q,K} &\equiv \left(\int_K \dd^3 k \abs{f(k)}^q\right)^{1/q} \intertext{and} 
   \forall f,g\in L^2(K),\ \avg{f,g}_K &\equiv \int_K \dd^3 k f(k)g^*(k) \mperiod
\end{align}
We often leave $K$ implicit when the result holds for any $K$. Note that for $0<q<1$, $\norm{\cdot}_q$  is not a norm, but we preserve the same notations provided the integral is finite. 

\begin{itemize}
    \item {\it Lemma.} Fix $C\in \C$. If $q\gs 1$, 
 let $s> 1$ be the H\"older conjugate of $q$, i.e. $1/q+ 1/s=1$ (for $q=1$, $s=\infty$ and $\norm{\cdot}_s = \norm{\cdot}_\infty$). 

 If $q\gs 1$,
\begin{align}
\label{eq:inf-pb-appx}
    \forall g\in L^2(K)\cap L^s(K), &\inf_{f\in L^2(K)\cap L^q(K)}\left\{ (\norm{f}_{q})^q \ | \ \avg{f,g} = C\right\} = \left(\frac{|C|}{\norm{g}_{s}}\right)^q \mperiod 
    \intertext{If $0<q< 1$,}\forall g\in L^2(K)\cap L^\infty(K), &\inf_{f\in L^2(K)\cap L^q(K)}\left\{ (\norm{f}_{q})^q \ | \ \avg{f,g} = C\right\} = \vbar^{q-1}\left(\frac{|C|}{\norm{g}_{\infty}}\right)^q \mperiod
\end{align}
 In the second case, we had to introduce $V\subsc{1cell} = 1/\vbar$, the measure (i.e. volume) occupied by singletons (i.e. one cell) in $K$ (i.e. in Fourier space). If the measure on $K$ is continuous (i.e. the Fourier transform is continuous), $1/\vbar = 0$ and the infimum is zero because $1-q>0$.
 
\item {\it Proof for $q\gs 1$. } We have $\abs{C} = \left|\avg{f,g}\right| \ls \norm{fg}_1$. By Hölder's inequality, $\norm{fg}_1 \ls \norm{f}_q \norm{g}_s ~\forall f\in L^q$, from which it follows that $$\inf_{f}\norm{f}_q^q \gs \abs{C}^q / \norm{g}_s^q\mperiod $$
Now we show that this lower bound is reached by some $f$.

Consider first $q>1$. Equality is reached if and only if $\exists \lambda\in \R_+$, $\abs{f}^q = \lambda \abs{g}^s$ almost everywhere, i.e. $\abs{f}=\lambda \abs{g}^{s-1}$ (as $s/q=s-1$). We can then fix $\lambda$ so that $\avg{f,g}=C$. Indeed, fix $f(k) = \lambda e^{i \arg(C)} e^{i \theta_g(k)} \abs{g(k)}^{s-1}$ where $\theta_g(k) = \arg(g(k))$. It satisfies the equality case of Hölder's inequality and $\avg{f,g}=\int f g^{*} = \lambda e^{i \arg(C)} \int \abs{g}^{s-1} e^{i(\theta_g(k)-\theta_g(k))}\abs{g^*} = \lambda e^{i \arg(C)} \norm{g}_s^s$.  We then choose $\lambda = \abs{C} / \norm{g}_s^s$ that is a positive real number, as required. We then have $\avg{f,g}=C$ and $\abs{\avg{f,g}} = \norm{fg}_1$.

Consider now $q=1$ ($s=\infty$). By simply writing $\norm{fg}_1 = \int \abs{f}\abs{g} \ls \int \abs{f} \norm{g}_\infty = \norm{f}_1 \norm{g}_\infty$, we see that the equality case imposes $\abs{g(k)} = \norm{g}_\infty$ almost everywhere on the support of $f$, $\{k| f(k)\neq 0\}$. Then, we choose $f$ to be zero wherever $g(k)\neq \norm{g}_\infty$, and else $f(k) = \lambda e^{i\arg(C) + \arg(g(k))}$  with $\lambda = \abs{C}/(\norm{g}_\infty \times\mu(\{g=\norm{g}_\infty\})>0$, where $\mu(A)$ denotes the measure of the set $A$.\footnote{For this proof to hold, $g(k)$ must therefore equates the value of $\norm{g}_\infty$ on a non-zero measure set. In our application, $g(k)=e^{ikx}$ and this is indeed the case.} We find $\avg{f,g}=C$ and $\abs{\avg{f,g}} = \norm{fg}_1$.
In both cases, for that particular $f$, $\norm{f}_q^q = \abs{C}^q / \norm{g}_s^q$. \hfill $\square$

\item {\it Corollary.} This must be slightly adapted to answer our problem \eqref{eq:full-expr-pdR} when the Fourier integral is only performed on the half-space $\Ralft$ or $\Kalf(R)$ (see Appendix \ref{appx:homo}). Determining the solution of \Eq\eqref{eq:full-expr-pdR} amounts to find
\begin{align}
    &\inf_f\{ \Vert f\Vert_{q,\Ralft}^q \ | \ 2 {\rm Re}(\avg{f,g}_{\Ralft}) =C\} \\
    =&\inf_f\{ \frac{1}{2}\Vert f\Vert_{q,\R^3}^q \ | \ \avg{f,g}_{\R^3} =C \ \text{and}\ f(-k) = f^*(k)\} \\
    \gs & \frac{1}{2} \inf_f\{\Vert f\Vert_{q,\R^3}^q \ | \ \avg{f,g}_{\R^3} =C \} = \frac{\abs{C}^q}{2\norm{g}_{s,\R^3}^q} \mperiod
\end{align}
However, one observes in the lemma's proof that the function $f$ built  such that $\Vert f\Vert^q_q = \abs{C}^q / (2\norm{g}_{s}^q)$ actually satisfies $f(-k)=f^*(k)$ when $C\in \R$ and $g(-k) = g^*(k)$ (recall that $g(k) = e^{i kx}$). As a consequence, the last inequality above is an equality.

\item {\it Proof for $0<q<1$.} Hölder inequality is lost in this case, because it relies on the convexity of unit $q$-balls. Let us start with the continuous measure case and show how the infimum is now zero. Suppose that we can find a sequence of subsets $A_\eps \subset K$ such that {\it (i)}  $\forall \eps >0$, $A_\eps$ has measure $\mu(A_\eps) = \eps$; {\it (ii)} $\exists m>0,\forall \eps >0, \inf_{A_\eps}\abs{g} \gs m$ (note that $m$ is independent of $\eps$). In our application $g(k) = e^{ikx}$, so $m=1$ works, and one can take $A_\eps$ to be the ball centered at $k=0$ of volume $\eps$. 

Fix $\eps >0$ and $f_\eps(k) = \lambda_\eps e^{i \arg(C)+i \th_g(k))} \mathbbm{1}_{A_\eps}(k)$ where $\lambda_\eps = \abs{C} / \int_{A_{\eps}} \abs{g}$ and $\mathbbm{1}$ is the indicator function. Then $\avg{f,g} = C$ and $\norm{f_\eps}_q^q = \int \abs{f}^q = \abs{\lambda_\eps}^q \times \mu(A_\eps) \ls \frac{\abs{C}^q \mu(A_\eps)}{m^q  \mu(A_\eps)^q} \to 0$ when $\eps\to 0$ because $\mu(A_\eps)\to 0$ and, crucially, $1-q>0$. Hence the infimum is zero.

However, this proof fails if one cannot build a sequence of sets with arbitrary small volumes, i.e. if the measure has atoms. If this is the case let us write $\int \dd^3 k = \vbar^{-1} \sum_k$ with $\sum_k$ a discrete finite sum and $\vbar^{-1}$ the measure of a singleton ($\vbar$ is a large volume in real space hence $\vbar^{-1}$ a small volume in Fourier space, see the notations in the beginning of this article). It is well-known that any concave function $h:\R_+\to\R_+$ with $h(0)=0$ is sub-additive, i.e.
\begin{equation}
    \forall x_1,\dots,x_n \gs 0,\ h\left(\sum_i x_i\right) \ls \sum_i h(x_i) \mperiod
\end{equation}
Although the inequality may seem to require convexity, this is indeed a concavity property. Applying this to $h: x \to x^q$ for $0<q<1$ we get
\begin{align}
    \norm{f}_q^q &= \vbar^{-1} \sum_k \abs{f}^q \\
    & \gs \vbar^{-1} \left(\sum_k \abs{f}\right)^q \mcomma
\end{align}
and if $\avg{f,g}=C$ we further have $\abs{C} \ls \vbar^{-1} \sum_k \abs{f} \abs{g} \ls \vbar^{-1} \norm{g}_\infty \sum_k \abs{f}$, so $\norm{f}_q^q \gs \vbar^{q-1}\abs{C}^q / \norm{g}^q_\infty$. Let $x_0$ be in $\{k\mid \abs{g(k)} = \norm{g}_\infty\}$. This lower bound on $\norm{f}_q^q$ is reached for $f(k) = \vbar C e^{i \arg(g)} \mathbbm{1}_{\{x_0\}}(k) / \norm{g}_\infty$, and is thus the infimum. It reproduces the case $q=1$ when $q\to 1^-$. It also goes to $0$ if $\vbar^{-1} \to 0$, reproducing the continuous measure case. \hfill $\square$

\end{itemize}

\section{Random walk and probabilities}
\label{appx:random-walks}

\subsection{Theorem bestiary}
\label{appx:sub:rw-theorems}

We list here some important, well-known theorems in probability and random walk theory that have been used to derive the expressions presented in the main text. We only sketch their formulation, and the reader interested in more formal statements may refer to Ref.~\cite{feller1971probability_book}. 

We particularly emphasize that none of these theorems require the underlying probability distributions to be Gaussian. This makes them particularly powerful for studying a wider class of distributions.  \bigskip

\begin{itemize}
    \item {\it Notations.} Let $\De:\R_+ \to \R$ be a random walk with initial condition $\De(t=0)=0$. We denote $S(t) = \sup_{0\ls u \ls t} \Delta(u)$. For $b>0$ (respectively $b<0$), the first-passage time at $b$ is denoted $T_b = \inf\{ t>0 | \De(t) \gs b\}$ (respectively $T_b = \inf_{t>0}\{t| \De(t)\ls b\}$). For $b>0$ the events $[S(t) \gs b]$ and $[T_b \ls t]$ are thus equal. 

    \item {\it Reflexion principle.} Assume that the random walk  {\it (i)} is symmetrical ($\De(t)$ and $-\De(t)$ share the same distribution) and {\it (ii)} satisfies the strong Markov property \cite{feller1971probability_book}. Then
\begin{equation}
    \forall t\gs 0,\forall~ 0<b<c,\hspace*{0.1cm}  \Pro(S(t) \gs c\ \text{and}\ \De(t) \ls b) = \Pro(\Delta(t) \gs 2c-b) \mperiod
\end{equation}

\item {\it Proof.} Although this proof can be found in the literature \cite{Bayraktar_2015_weak_reflection_principle}, we reproduce it to convince the reader that Gaussianity is not required. Define the shifted random motion $\De'(t) = \De(t+T_c) - \De(T_c)$. By definition, $\De(T_c) = c$. Then 
\begin{align}
    \Pro(S(t) \gs c, \De(t) \ls b) &= \Pro(T_c \ls t, \De(t-T_c + T_c) \ls b) \\
    & =  \Pro(T_c \ls t, \De'(t-T_c) \ls b - \De(T_c)) \\
    &=\Pro(T_c \ls t, \De'(t-T_c) \ls b - c) \\
    &= \Pro(T_c \ls t, \De'(t-T_c) \gs c-b) \\
    &=\Pro(T_c \ls t, \De(t) \gs 2c-b) \\
    &= \Pro(\De(t) \gs 2c-b) \quad \text{since $2c-b > c$.}  
\end{align}
In the fourth equality we have used combined the strong Markov property ensuring that $\De'$ has the same distribution as $\De$, and the assumption that $\De$ and $-\De$ have the same distribution.\hfill $\square$

\item {\it Corollary.} $S(t)$ and $|\De(t)|$ are identically distributed. Indeed, $\forall b>0$, $\Pro(S(t) \gs b) = \Pro(S(t) \gs b, \De(t) \gs b) + \Pro(S(t) \gs b, \De(t) \ls b) = \Pro(\De(t) \gs b) + \Pro(\De(t) \gs 2b -b) = \Pro(\abs{\De(t)} \gs b)$, where the second equality stems from the reflexion principle. 

Note however that in the space of all possible random paths, the paths $t\mapsto S(t)$ and $t\mapsto \abs{\De(t)}$ are not identically distributed (since, e.g, $S$ must be monotonic whereas $\abs{\De}$ is not necessarily so).

\item {\it First-passage time distribution.} Assume that {\it (i)} the hypotheses of the reflexion principle are fulfilled, and {\it (ii)} the random walk satisfies a ``diffusion property"\footnote{This is usually verified when the variance of the walk is $\avg{\Delta^2(t)} \propto t$. It applies to the class of probabilities \eqref{eq:pdt-with-polynom}.}, namely
\begin{equation}
    \forall t>0,\forall b>0,~ \Pro(|\De(t)|\gs b) = \Pro(\sqrt{t} |\De(1)| \gs b) \mperiod
\end{equation}
Then, $T_b$ has the same distribution as $b^2 / \De(1)^2$.

\item {\it Proof.} Using the previous corollary, $\forall t>0,\forall b>0, \ \Pro(T_b \ls t) = \Pro(S(t) \gs b) = \Pro(\abs{\De(t)} \gs b) = \Pro(\sqrt{t}\abs{\De(1)} \gs b) = \Pro(b^2/\De(1)^2 \ls t)$, so the cumulative distribution functions of $T_b$ and $b^2/\De(1)^2$ are identical.

\medskip

\item {\it Corollary.} It follows from the last property and the transfer theorem that the probability density of the first-passage time at $b>0$ is given by
\begin{equation}
    p_{T_b}(t) = \frac{b}{t^{3/2}} \times p_{\De(1)}\left(\frac{b}{\sqrt{t}}\right)\mperiod
\end{equation}
This last equation, providing a direct link between the distribution of the random walk and that of the one-barrier first-passage time, is \Eq \eqref{eq:1st-passage-time-vs-walk} in the main text. By symmetry, this also applies to negative barriers by replacing $b$ by $\abs{b}$ above.

\end{itemize}

\subsection{Solution to the two-barrier problem}
\label{appx:sub:2-barriers}

In the study of voids using the excursion set formalism, an additional complication is the existence of the ``void-in-cloud" effect, which is modeled by the presence of {\it two} barriers \cite{Sheth:2003py_seminal}. 

Fix barriers $a<0<b$ and $c$ be of either sign.  In the following, $p_1(t,c)$ refers to the probability of the first-passage time at barrier $c$, i.e., $p_1(t,c)\dd t=p_{T_c}(t) \dd t = \Pro(T_c \in \interval{[}{t}{t+\dd t}{[})$. We similarly denote by $p_2(t,a,b)$  the probability density of the first-passage time at barrier $a<0$ happening before the passage at $b>0$, i.e., $p_2(t,a,b)\dd t=\Pro(T_a \in \interval{[}{t}{t+\dd t}{[}\  \cap\ T_a < T_b) $. Their respective Laplace transforms $\ell_1$ and $\ell_2$ are defined as
\begin{equation}
    \ell_1(s,c) \equiv \int_0^\infty e^{-st} p_1(t,c)\dd t\ , \qquad \text{and similarly for $p_2$ and $\ell_2$.}
\end{equation}
The authors of Ref.~\cite{Sheth:2003py_seminal} have shown the following relation between these Laplace transforms:
\begin{equation}
\label{eq:1-to-2-barriers}
    \ell_2(s,a,b) = \frac{\ell_1(s,a)-\ell_1(s,b)\ell_1(s,b-a)}{1-\ell_1(s,b-a)^2}\mcomma
\end{equation}
where the symmetry property, $p_1(t,-c)=p_1(t,c)$ is used by Ref. ~\cite{Sheth:2003py_seminal} to derive \eqref{eq:1-to-2-barriers}. We now consider the family of probability densities for the single barrier problem established in \Eq \eqref{eq:1st-passage-expo-tail},
\begin{equation}
\label{eq:1bar-passage-time-appx}
    p_1(t,c) = \frac{q}{2\Gamma((1+\al)/q)} \frac{\abs{c}\gamma}{t^{3/2}}\left(\frac{\abs{c} \gamma}{\sqrt{t}}\right)^\al \exp[- (\abs{c} \gamma/\sqrt{t})^q]
\end{equation}
where $q>0,\al>-1$ and $\gamma \equiv \left(\Gamma((3+\al)/q)/\Gamma((1+\al)/q)\right)^{1/2}$.  We recall that the exponent $\al$ is a degree of freedom, as the large deviation principle is insensitive to polynomial corrections to the probability tail.

Only in the Gaussian case $q=2$ does $\ell_1(s,a)$ have a simple analytical form. However, we can derive its behavior for $s\to+\infty$ in the general case using Laplace's method for approximations of integrals. A few changes of variables allow us to recast $\ell_1(s,c)$ as
\begin{equation}\label{eq:exact-laplace}
    \ell_1(s,c) = \frac{q}{\Gamma((1+\al)/q)}(s (c\gamma)^2)^{1/(q+2)}\int_0^\infty \dd u \left[ u (s (c\gamma)^2)^{1/(q+2)}\right]^\al \exp[-\lambda\left( \frac{1}{u^2}+u^q\right)]\mperiod
\end{equation}
with parameter 
\begin{equation}
    \lambda \equiv (s(c\gamma)^2)^{q/(q+2)} \xrightarrow[s\to+\infty]{}+\infty \mperiod
\end{equation}
Performing Laplace's method then leads to 
\begin{equation}
\label{eq:complicated-laplace}
    \ell_1(s,c) \underset{s\to +\infty}{\simeq} \frac{ \sqrt{\pi Q}}{\Gamma((1+\al)/q)} \xi_{s,c}^{(2-q)/(2(q+2))} \xi_{s,c}^{\al/(q+2)} \exp\left(- \frac{q+2}{2} \xi_{s,c}^{Q/2}\right)\mcomma
\end{equation}
with 
\begin{equation}
    Q\equiv \frac{2q}{q+2} >0\spliteq \xi_{s,c}\equiv \frac{2 s (c\gamma)^2}{q}  \mperiod
\end{equation}
Note that $Q=1$ for the Gaussian case $q=2$, and increases with increasing $q$. In order to simplify \Eq \eqref{eq:1-to-2-barriers}, we now fully exploit the wiggle room provided by the large deviation principle by picking a convenient exponent $\al$. Choosing
\begin{equation}
    \al = \frac{q-2}{2}>-1\mcomma
\end{equation}
the expression \eqref{eq:complicated-laplace} simplifies to
\begin{equation}
\label{eq:simple-laplace}
    \mathrm{L.T.}[p_1(t,c)](s) =\ell_1(s,c) \underset{s\to +\infty}{\simeq} \sqrt{Q}\exp(- \frac{q+2}{2} \xi_{s,c}^{Q/2}) \mcomma
\end{equation}
with now $\gamma = (\Gamma(1/2 + 2/q)/\sqrt{\pi})^{1/2}$. This reproduces the result of Ref.~\cite{Sheth:2003py_seminal} for $q=2$, which is exact for all $s$ in this case. Expanding the denominator of \Eq\eqref{eq:1-to-2-barriers}  in series\footnote{A sufficient condition to be able to do this is $Q\ls 1 $ i.e. $0<q\ls 2$. In practice because $  s\gg 1$ the exponential is very small, and the validity holds for any $q>0$.}, we obtain 
\begin{align}
\nonumber
    \ell_2(s,a,b) &\underset{s\to +\infty}{\simeq} \sum_{n\gs 0}\left[Q^{n+1/2} \exp\left[-\frac{q+2}{2}\left(\frac{2s\gamma^2}{q}\right)^{Q/2}(\abs{a}^{Q}+2n(b-a)^{Q})\right]\right. \\
    \label{eq:Laplace-infinite-sum}
    &\quad\left.-\ Q^{n+1}\exp\left[-\frac{q+2}{2}\left(\frac{2s\gamma^2}{q}\right)^{Q/2}(b^{Q}+(2n+1)(b-a)^{Q})\right]\right] \mperiod
\end{align}

\begin{figure}
    \centering
    \includegraphics[width=\linewidth]{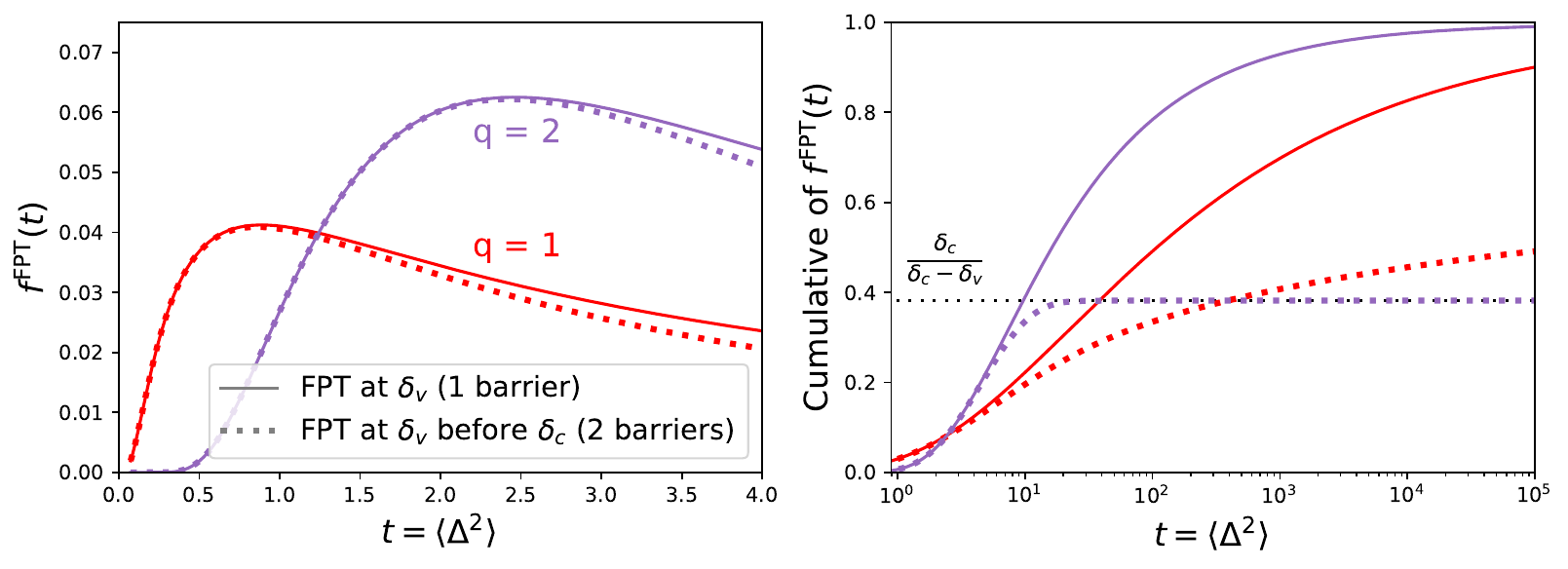}
    \caption{{\it Left:} Comparison of the FPT distributions \eqref{eq:1bar-passage-time-appx} at $c=a=\delta_v$ (one barrier, solid curves) and \eqref{eq:fv-appx} at $a=\delta_v, b=\delta_c$ (two barriers, dotted curves). As expected, they become undistinguishable in the very small $t$ limit. {\it Right:} Cumulative distributions $F(t) = \int_0^t f\fpt(t')\dd t'$ of the functions illustrated on the left. Both solid curves converge to $1$ when $t\to+\infty$. A contrario, the Gaussian double barrier distribution converges to $F(t)\to\delta_c /(\delta_c -\delta_v)$, according to the gambler's ruin principle \cite{Sheth:2003py_seminal}. For $q\neq 2$ however the two-barrier distribution does not converge to the same limit, illustrating that expressions derived with the LDP are only valid in the limit $t < \delta_c, \abs{\delta_v}$.}
    \label{fig:2bar-vs-1bar}
\end{figure}

We now observe that even though this expression is not directly invertible, we can find a probability density whose Laplace transform evaluated in the limit $s\to+\infty$  with Laplace's method is precisely \Eq \eqref{eq:Laplace-infinite-sum}. We first define new barriers $ \forall n\gs0$,
\begin{equation}
    c_n \equiv (\abs{a}^{Q}+2n(b-a)^{Q})^{1/Q}\quad,\quad d_n \equiv (b^{Q}+(2n+1)(b-a)^{Q})^{1/Q}\mperiod
\end{equation}
The link we have established between \Eqs\eqref{eq:1bar-passage-time-appx} and \eqref{eq:simple-laplace} then leads to
\begin{align}
\nonumber
     p_2(t,a,b)  &= \frac{q}{2\sqrt{\pi} t} \sum_{n\gs 0}\left[Q^{n} \left(\frac{c_n \gamma}{\sqrt{t}}\right)^{q/2}\exp\left(-\left(\frac{c_n \gamma}{\sqrt{t}}\right)^{q}\right)\right. \\
     \label{eq:fv-appx}
    &\quad\left.-\ Q^{n+1/2}\left(\frac{d_n \gamma}{\sqrt{t}}\right)^{q/2}\exp\left(-\left(\frac{d_n \gamma}{\sqrt{t}}\right)^{q}\right)\right]\mperiod
\end{align}

This is \Eq\eqref{eq:fv-main-text} in the main text. In the Gaussian case $q=2$, $\gamma=1/\sqrt{2}$ and we recover the equation (8) of Ref.~\cite{DAmico:2010dwy_void_png}. Let us remark that by Gambler's ruin property, $p_2$ should in principle be normalized to $\int_0^\infty p_2(t,a,b)\dd t = b/(b-a)$. This is however not easy to check, as the integral and sum cannot be exchanged (this problem is already present in Eq. (8) of Ref.~ \cite{DAmico:2010dwy_void_png}). One should rather use $\int_0^\infty p_2(t,a,b)\dd t =\ell_2(0,a,b)$ and take the limit $s\to 0$ of \Eq\eqref{eq:1-to-2-barriers}. However, the integral \eqref{eq:exact-laplace} cannot be solved analytically for general $q$ and our expression  \eqref{eq:simple-laplace} for $\ell_1$ has been derived in the large $s$ limit, so cannot be used for $s=0$. In the Gaussian case, $q=2$, the approximation \eqref{eq:simple-laplace} is actually exact and inserting it in expression \eqref{eq:1-to-2-barriers} one finds the correct normalization. Care is required in taking limits. One has to set $q=2$ first, then take the limit $s\to 0$. In \fig\ref{fig:2bar-vs-1bar} we show these normalizations and compare $p_2$ to $p_1$ in order to validate some consistency checks.

\bibliographystyle{JHEP}
\bibliography{references}

\end{document}